\documentclass[acus]{JAC2003}

\usepackage{graphicx,epsfig}

\newcommand{\ra}{\rightarrow}
\newcommand{\ee}{{\rm e^+e^-}}
\newcommand{\ffbar}{{\rm f\bar f}}
\newcommand{\bb}{{\rm b\bar b}}
\newcommand{\cc}{{\rm c\bar c}}
\newcommand{\ttbar}{{\rm t\bar t}}
\newcommand{\tautau}{{\tau^+ \tau^-}}
\newcommand{\glgl}{{\rm gg}}

\newcommand{\nenebar}{{\nu_e \bar \nu_e}}

\newcommand{\qq}{{\rm q \bar q}}
\newcommand{\mumu}{\mu^+\mu^-}
\newcommand{\ellell}{\ell^+ \ell^-}
\newcommand{\lnu}{\ell^-\bar\nu}
\newcommand{\WW}{{\rm W^+ W^-}}

\newcommand{\gaga}{\gamma\gamma}
\newcommand{\Z}{{\rm Z}}

\newcommand{\Hp}{{\rm H^+}}
\newcommand{\Hm}{{\rm H^-}}
\newcommand{\Hpm}{{\rm H^\pm}}
\newcommand{\lh}{{\rm h^0}}
\newcommand{\hh}{{\rm H^0}}
\newcommand{\Ho}{{\rm H^0}}
\newcommand{\Hosm}{{\rm H^0_{\mathrm{SM}}}}
\newcommand{\Ao}{{\rm A^0}}
\newcommand{\Hi}{\ensuremath{\mathrm{H}_i}}
\newcommand{\Hj}{\ensuremath{\mathrm{H}_j}}

\newcommand{\tb}{{\tan\beta}}

\newcommand{\mH}{{\rm m}_{\rm H}}
\newcommand{\mlh}{{\rm m}_{\rm h}}
\newcommand{\mA}{{\rm m}_{\rm A}}
\newcommand{\mHpm}{{\rm m}_{{\rm H}^\pm}}
\newcommand{\mZ}{{\rm m}_{\rm Z}}
\newcommand{\mW}{{\rm m}_{\rm W}}
\newcommand{\mt}{{\rm m}_{\rm t}}
\newcommand{\fb}{{\rm fb}^{-1}}
\newcommand{\fbc}{{\rm fb}}
\newcommand{\ab}{{\rm ab}^{-1}}
\newcommand{\sqrts}{{\sqrt{s}}}

\begin{document}
\title{HIGGS BOSON PRECISION STUDIES AT A LINEAR COLLIDER
\thanks{
Most of the work reported in this talk was done by members
of the Higgs working group of the Extended ECFA/DESY Study: 
V.~Barger$^{a}$, M.~Battaglia$^{b}$, M.~Beccaria$^{c}$, 
E.~Boos$^{d}$, J.C.~Brient$^e$, 
S.Y.~Choi$^{f}$, D.~Choudhury$^{b}$,
A.~Datta$^{g}$, S.~Dawson$^{h}$, S.~DeCurtis$^{i}$, 
G.~Degrassi$^{j,k}$, A.~Denner$^{l}$, A.~DeRoeck$^{b}$, 
N.G.~Deshpande$^{m}$, S.~Dittmaier$^{n}$, 
A.~Djouadi$^{o}$, D.~Dominici$^{i,p}$, M.~Dubinin,$^{d}$
H.~Eberl$^{q}$, J.~Ellis$^{b}$, A.~Ferrari$^{r}$, M.~Frank$^{s}$, 
E.~Gabrielli$^{g}$, A.~Gay$^{t}$, I.F.~Ginzburg$^{u}$, 
D.K.~Ghosh$^{m}$, E.~Gross$^{v}$, J.~Guasch$^{l}$, J.F.~Gunion$^{w}$, 
T.~Hahn$^{n}$, T.~Han$^{a}$,S.~Heinemeyer$^{x}$, W.~Hollik$^{n}$, 
K.~Huitu$^{g}$,
A.~Imhof$^{y,z}$, J.~Jiang$^{aa}$, A.~Kiiskinen$^{g}$, 
T.~Klimkovich$^{y,z}$, B.A.~Kniehl$^{z}$, M.~Krawczyk$^{bb}$, T.~Kuhl$^{y}$ 
P.~Langacker$^{cc}$, 
F.~Madricardo $^{z}$, W.~Majerotto$^{q}$, T.~Maki$^{g}$, B.~McElrath$^{a}$,
B.~Mele$^{j,k}$, N.~Meyer$^{y,z}$, D.J.~Miller$^{b}$, 
S.~Moretti$^{b}$, M.~M\"uhlleitner$^{l}$, 
K.~Olive$^{dd}$, P.~Osland$^{ee}$, 
S.~Pe\~{n}aranda$^{n}$, A.~Pilaftsis$^{ff}$,
A.~Raspereza$^{y}$, F.M.~Renard$^{o}$, 
M.~Ronan$^{gg}$, M.~Roth$^{n}$,
H.J.~Schreiber$^{y}$, 
M.~Schumacher$^{hh}$,
P.~Slavich$^{n,s}$, A.~Sopczak$^{ii}$, 
V.C.~Spanos$^{dd}$, M.~Steinhauser$^{y}$,
S.~Trimarchi$^{jj}$,
C.~Verzegnassi$^{jj}$, A.~Vologdin$^{d}$, 
Z.~Was$^{kk}$, M.M.~Weber$^{l}$, G.~Weiglein$^{ll}$, M.~Worek$^{mm}$
M.~Yao$^{gg}$, 
P.M.~Zerwas$^{y}$,
$^a$ University of Wisconsin, 
$^b$ CERN,
$^c$ INFN, University di Lecce,
$^d$ Moscow State University,
$^e$ LPNHE Ecole Polytechnique ,
$^f$ Chonbuk National University,
$^g$ Helsinki Instiute of Physics ,  
$^h$ BNL Brookhaven National Laboratory,
$^i$ INFN, Firenze, 
$^j$ INFN, Roma, 
$^k$ University La Sapienza, Roma,
$^l$ PSI Villigen,
$^m$ University of Oregon
$^n$ MPI M\"unchen,
$^o$ Universite Montpellier,
$^p$ University of Florence, 
$^q$ Inst.f.Hochenergiephysik Oesterr.Akademie d.Wissenschaften, Wien,
$^r$ Uppsala University,
$^s$ University Karlsruhe,
$^t$ IRES Strasbourg, 
$^u$ NSC Novosibirsk,
$^v$ Weizmann Institute, 
$^w$ University of California, Davis,
$^x$ LMU M\"unchen,
$^y$ DESY Hamburg,
$^z$ University of Hamburg,
$^{aa}$ ANL Argonne, 
$^{bb}$ Warsaw University,  
$^{cc}$ University of Pennsylvania, 
$^{dd}$ University of Minnesota,
$^{ee}$ University of Bergen, 
$^{ff}$ Manchester University,
$^{gg}$ LBNL Berkeley,
$^{hh}$ University of Bonn,
$^{ii}$ Lancaster University, 
$^{jj}$ INFN, University Trieste, 
$^{kk}$ INP Cracow,
$^{ll}$ University of Durham,
$^{mm}$ University of Silesia, Katowice.
}}

\author{Klaus Desch, University of Hamburg, Germany}

\maketitle

\begin{abstract}
This report summarizes the progress in the study of Higgs physics at
a future linear electron positron collider at center-of-mass energies
up to about 1000~GeV and high luminosity. 
After the publication of the TESLA Technical Design Report~\cite{TDR}, an extended
ECFA/DESY study on linear collider physics and detectors was performed.
The paper summarizes the status of the studies 
with main emphasis on recent results obtained in the course of the workshop.
\end{abstract}

\section{OBJECTIVES OF THE STUDY}

Elucidating the mechanism responsible for electro-weak symmetry
breaking is one of the most important tasks of future collider
based particle physics.
Experimental and theoretical indications of a light Higgs boson make
the precision study of the properties of Higgs bosons one of the
major physics motivations of a linear collider (LC).
Both the Higgs boson of the Standard Model (SM) and those of extended models will be
copiously produced in $\ee$ collisions in various production mechanisms.
A large variety of different decay modes can be observed with low backgrounds
and high efficiency. These measurements allow us to extract the fundamental 
parameters of the Higgs sector with high precision. 
The series of ECFA/DESY workshops aims at a comprehensive study of the
physics case, a determination of the achievable precisions on Higgs
observables as well as on a fruitful cross-talk between theory,
physics simulations and detector layout.

A future linear collider offers also the option of photon-photon collisions
from back-scattered laser light. The physics potential and progress in
Higgs physics at a photon collider is discussed elsewhere in these
proceedings~\cite{albert}.

\section{STANDARD MODEL HIGGS BOSON}

\subsection{Theoretical Predictions}
\begin{figure}[htb]
\centering
\epsfig{width=0.47\linewidth,file=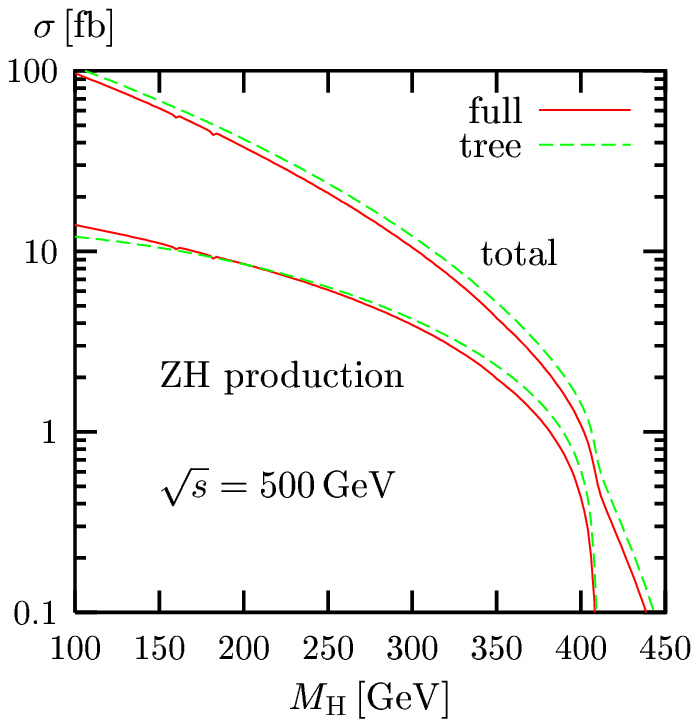}
\epsfig{width=0.47\linewidth,file=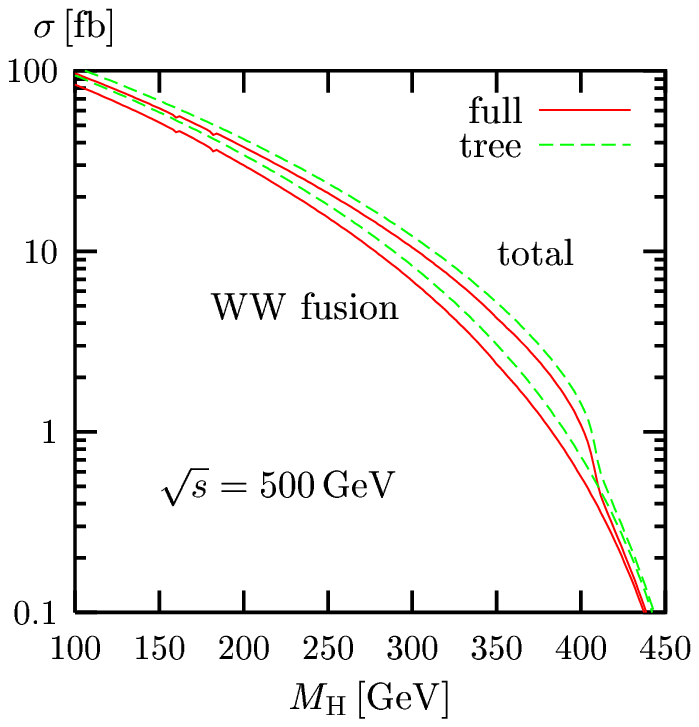}\\
\epsfig{width=0.47\linewidth,file=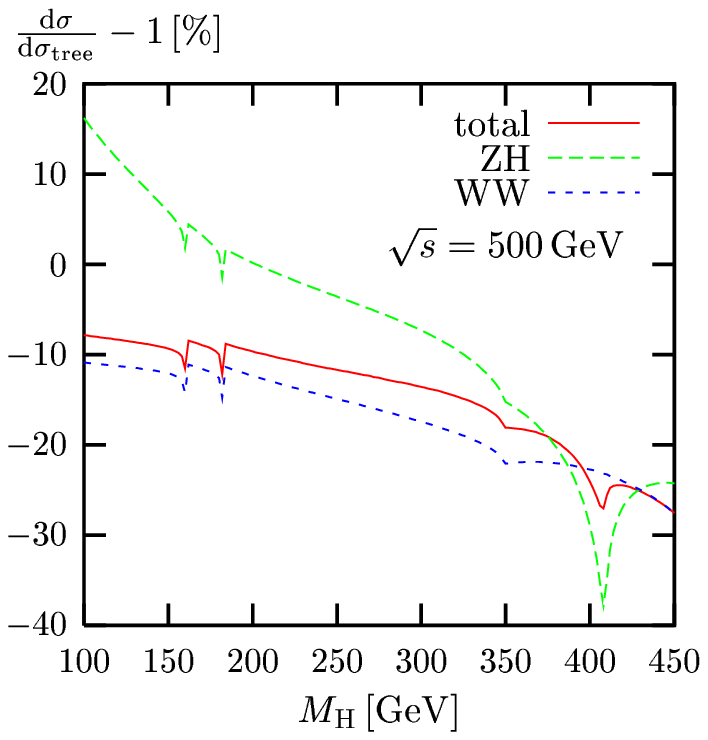}
\epsfig{width=0.47\linewidth,file=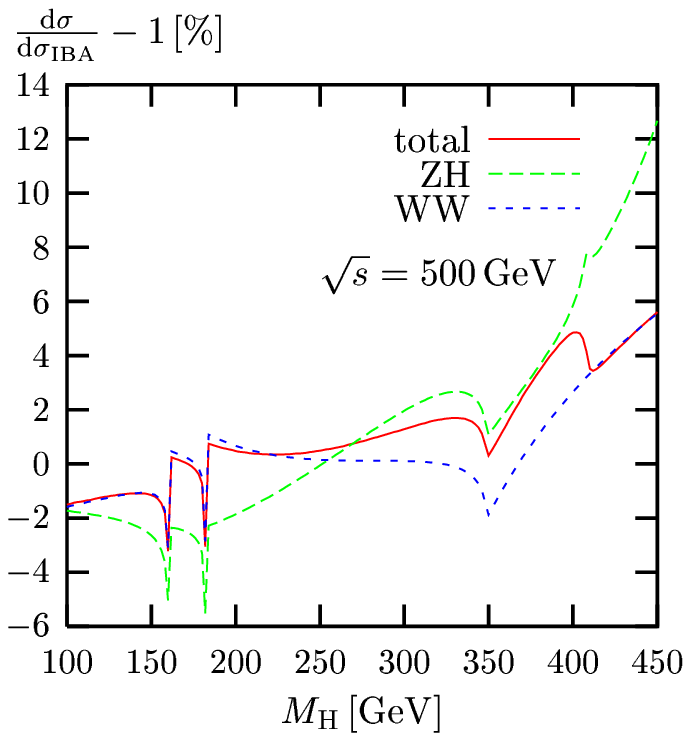}

\caption{Upper plots: cross-section for the processes $\ee\to\Z\Ho$ and 
$\ee\to\nenebar\Ho$ including complete one-loop electro-weak corrections for 
$\sqrt{s} = 500 $ GeV. Lower plots: Relative amount of one-loop corrections
relative to Born level result (left) and relative to an improved Born approximation
(IBA) (from~\cite{LC-TH-2003-008}).} 
\label{fig:zhxsec}
\end{figure}

In $\ee$ collisions, the SM Higgs boson is predominantly produced through
the Higgs-strahlung process, $\ee\ra\Ho\Z$~\cite{hstrahl}
 and through the vector boson fusion
processes $\ee\ra\nenebar(\ee)\Ho$~\cite{fusion}. The SM production cross-sections
are precisely known including full electro-weak corrections at the
one-loop level. For a recent review of the theoretical calculations see
e.g.~\cite{Kniehl:2002id}. 
Recently the full one-loop corrections to the WW-fusion
process have been calculated \cite{LC-TH-2003-007,LC-TH-2003-008}. 
The radiatively corrected cross-sections for Higgs-strahlung and WW-fusion
are shown in Fig.~\ref{fig:zhxsec}. For Higgs-strahlung the corrections
are positive for small Higgs masses and negative for large Higgs masses and
are of ${\cal O} (10\%)$. 
For WW-fusion the corrections are of similar size but always negative.

\begin{figure}[htb]
\centering
\epsfig{width=1.1\linewidth,file=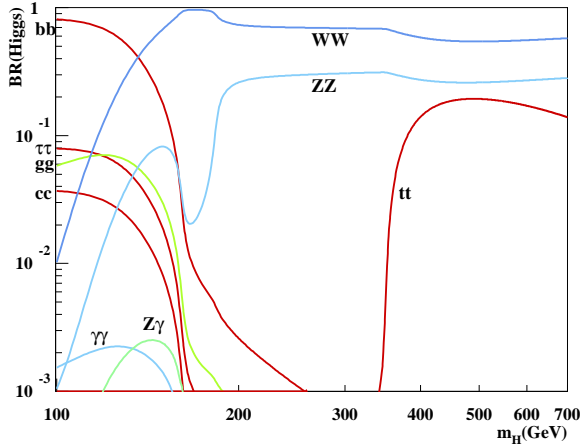}
\caption{Branching ratios of the SM Higgs boson calculated with
HDECAY~\cite{HDECAY}.}
\label{fig:hbr}
\end{figure}

With the Higgs boson being responsible for mass generation its couplings
to massive SM particles are proportional to their masses: 
$g_{ffH} = m_f / v$, $g_{VVH} = 2M_V^2/v$.
Thus Higgs bosons decay preferentially into the heaviest 
kinematically possible final states. 
State-of-the-art branching ratio calculations including 
electro-weak and QCD corrections~\cite{QCDDecay}
are coded in the program
HDECAY~\cite{HDECAY} for the SM and its minimal supersymmetric extension, the
MSSM. Branching ratios of the neutral Higgs bosons in the MSSM can be
also calculated with program FeynHiggsDecay~\cite{fhd}. 
The SM Higgs branching ratios in the mass range
relevant to a LC are shown in Fig.~\ref{fig:hbr}.

\subsection{Tools for Simulation}

A variety of leading-order Monte Carlo generators exist which are commonly
used for Higgs studies in $\ee $ collisions. They are PYTHIA~\cite{pythia},
HERWIG~\cite{herwig},
HZHA~\cite{hzha}, CompHep~\cite{comphep}, and WHiZard~\cite{whizard}.
CompHep and WHiZard offer the possibility of generating the complete
$2\to4$ and (in the case of WHiZard) also $2\to6$ processes including 
their interference with SM backgrounds.

Beamstrahlung was simulated in most analyses presented below using the 
parameterization CIRCE~\cite{CIRCE}.

The vast majority of experimental analyses in this summary includes the
simulation of complete SM backgrounds. The effects of limited detector
acceptance and resolution have been incorporated using the parametric
detector simulation program SIMDET~\cite{SIMDET} which is based on the
detector performance specified for the TESLA detector in the TDR. 
A comparative study of different event generators and of different
fast detector simulation programs was carried out in~\cite{ronan}.

Most analyses which involve tagging of heavy quarks use a realistic
event-wise neural-net algorithm based on 
ZVTOP~\cite{ZVTOP} which was first used at the SLD detector.

A detailed simulation (BRAHMS~\cite{BRAHMS}) of the TESLA TDR detector 
based on GEANT3 along 
with a reconstruction program is available and can be used for comparative
studies. 

\subsection{Coupling to Z Bosons}

\begin{figure}[htb]
\centering
\epsfig{height=0.92\linewidth,file=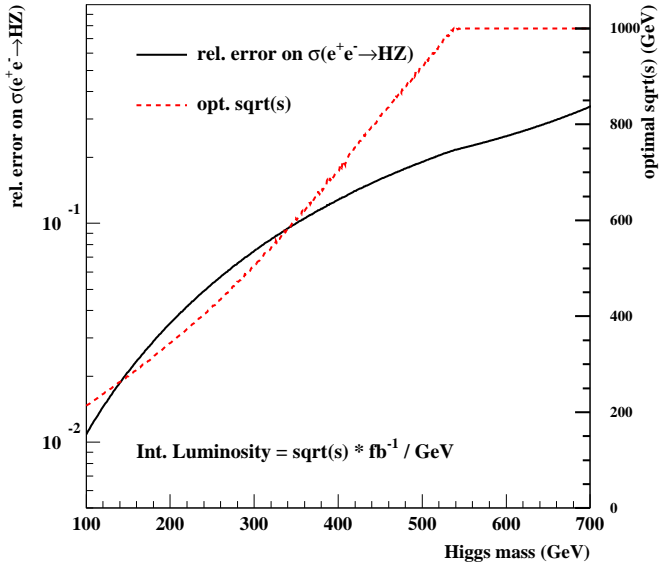}
\caption{Achievable precision on the cross-section for $\ee\ra\Ho\Z$ 
as a function of the Higgs mass. An integrated luminosity proportional
to the center-of-mass energy in fb$^{-1}$/GeV
at a center-of-mass energy corresponding to the maximum of the
cross-section is assumed. The center-of-mass energy which yields the 
largest cross-section is also shown (dashed line, right scale).}
\label{fig:recoil}
\end{figure}

The anchor of a model-independent precision analysis of Higgs boson
properties at a LC is the measurement of the total cross-section
for the Higgs-strahlung process, $\ee\ra\Ho\Z$. Z bosons can be selected
in $\Z\to\ee$ and $\Z\to\mumu$ decays. From energy-momentum
conservation the invariant mass recoiling against the $\Z$ candidate
can be calculated. Through a cut on the recoil mass,
Higgs bosons can be selected independent
of their decay mode, allowing for a model-independent measurement
of the effective HZ coupling, $g_{HZZ}$. 
Once $g_{HZZ}$ is known, all other Higgs couplings can
be determined absolutely. The total Higgs-strahlung cross-section can
be measured with an accuracy of 2.5\% for $\mH = 120 $ GeV and 
$\sqrts = 350 $ GeV for 500~fb$^{-1}$~\cite{lohmannhz}.
Assuming that the uncertainty scales
with the square root of the 
cross-section and that the selection purity and efficiency
is independent of the center-of-mass energy,
one can obtain an accuracy between 1.2 \% and 10\% 
for $100 < \mH < 360 $ GeV, for an integrated luminosity of 
$\sqrt{s} \times $ fb$^{-1}/$ GeV
at a center-of-mass energy corresponding to the maximum of the
cross-section for a given Higgs mass. The relative error
is shown in Fig.~\ref{fig:recoil} together with the optimal
center-of-mass energy as a function of the Higgs mass.

The importance of a precise and model-independent determination of
$g_{HZZ}$ has e.g.~recently been discussed in the context of supersymmetric
models~\cite{LC-TH-2003-005}
and in the context of models
with higher Higgs field representations, as well as in the context
of extra-dimensional models~\cite{huitu}.
\subsection{Quantum Numbers}

\begin{figure}[htb]
\centering
\epsfig{height=0.92\linewidth,file=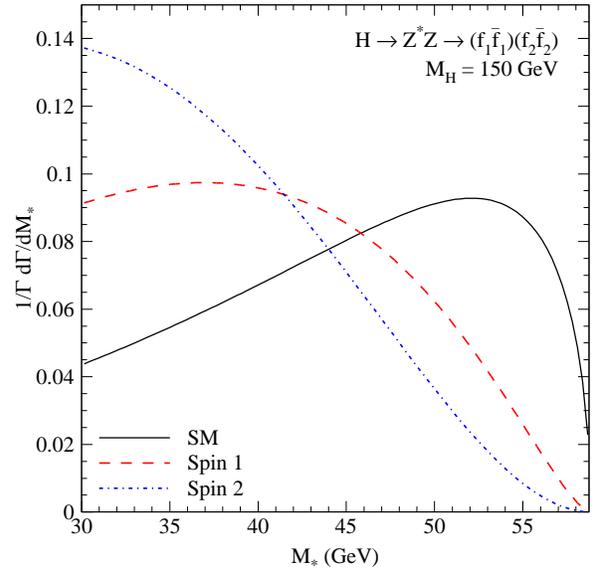}
\caption{Distribution of the of the invariant mass of the decay products of
the off-shell $\Z^*$ boson in $\Ho\ra\Z\Z^*$ decays for the SM Higgs and
for examples of spin-1 and spin-2 bosons for $\mH = 150 $ GeV
(from~\cite{LC-TH-2003-036}).}
\label{fig:hzz}
\end{figure}

The measurements of differential production cross-sections 
and decay angular distributions provide access
to the discrete quantum numbers of the Higgs boson: $J^{PC}$. 
In the TDR, the measurement of the $\beta$-dependence
of the Higgs-strahlung cross-section close to the production threshold 
was exploited to determine the spin of the Higgs boson.
The spin can also
be determined from the invariant mass of the off-shell $\Z$ boson in the
decay $\Ho\ra\Z\Z^*$ for $\mH < 2 \mZ$. 
This method is independent of the Higgs production process
and thus potentially applicable also in $\gamma\gamma$ and gg collisions.
The invariant mass distribution for $\mH = 150 $~GeV is shown in 
Fig.~\ref{fig:hzz}. For $\mH$ above $2\mZ$, azimuthal correlations of the two
Z boson decay planes can be exploited to gain 
sensitivity to Higgs boson spin and CP~\cite{Barger, LC-TH-2003-036}.

\begin{figure}[htb]
\centering
\epsfig{width=0.9\linewidth,file=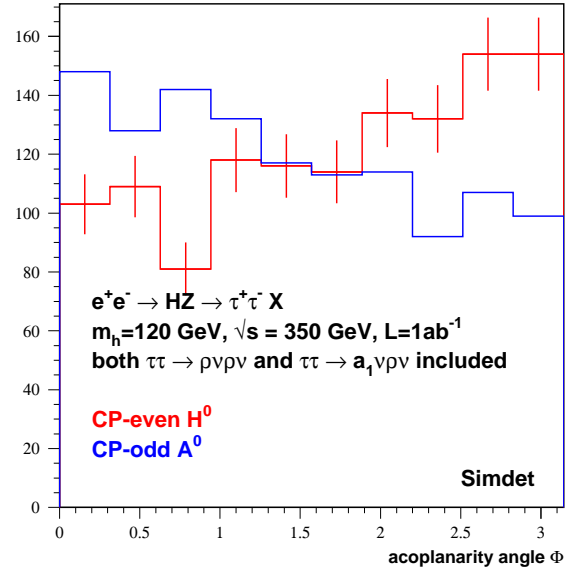}
\caption{Acoplanarity angle between the two $\rho$ decay planes from
$\Ho/\Ao\to\tautau$ 
decays (from~\cite{Imhof-Amsterdam}).}
\label{fig:andreas}
\end{figure}

The CP quantum number, like the spin, can be determined from both
Higgs boson production and decay~\cite{CP}. 
In the TDR, the sensitivity of the angular
distribution of the $\Z$ recoiling against the $\Ho$  in Higgs-strahlung was exploited. Recently a method has been proposed which makes use of the transverse
spin correlation in $\Ho\to\tau^+\tau^-$ decays. The spin correlations between
the two $\tau$ leptons is probed through angular correlations of their decay
products. In particular, events from $\tau^\pm\to\rho^\pm\nu_\tau\to\pi^\pm\pi^0
\nu_\tau$ and from $\tau^\pm\to a_1^\pm\nu_\tau\to\rho^0\pi^\pm\nu_\tau\to
\pi^\pm\pi^\mp\pi^\pm\nu_\tau$
can be used. The angle between the decay planes of the two $\rho$ mesons from
either $\tau$ decay provides a suitable observable~\cite{LC-PHSM-2003-050,
LC-PHSM-2003-049}. While this angle can be determined in the laboratory frame,
ideally it is evaluated in the Higgs boson rest frame, which can be 
approximately reconstructed using $\tau$ lifetime 
information~\cite{LC-PHSM-2003-003}. Preliminary results including detector
simulation have shown that from a sample of 1 ab$^{-1}$ of Higgs-strahlung 
events at $\sqrt{s} = 350 $~GeV, a statistical separation between a
CP-even and a CP-odd Higgs boson of eight standard deviations may be
achieved assuming production cross section and branching ratio 
as for $\Hosm$ (see Fig.~\ref{fig:andreas},
note that background is not yet taken into account)~\cite{Imhof-Amsterdam}.

\subsection{Decay Branching Ratios}

The precise measurement of Higgs boson decay branching ratios is one of
the key tasks in LC Higgs physics. In the TESLA TDR as well as in all other 
regional LC studies~\cite{snowmass,acfa} analyses have been performed to
investigate
the expected precisions on the branching ratio determination. For a light
Higgs boson with $\mH < 160 $~GeV, a large variety of Higgs decay modes
can be measured. The hadronic decays into $\bb, \cc, $ and $\glgl$ are
disentangled via the excellent capabilities of a LC vertex detector. 
Progress has been achieved recently in the level of detail at which the
algorithms to tag b- and c-quarks are implemented into the simulation. 
Although these studies are not finished, it looks conceivable that the
results of the TDR study will essentially be confirmed~\cite{kuhl}.

There are two different methods to extract branching ratios from the observed
events: 
\begin{enumerate}
\item Measure the topological cross-section for a given final state,
      e.g.~$\sigma(\Ho\Z\to X\Z)$ and divide by the total measured
      Higgs--strahlung cross-section (as obtained from the recoil mass
      measurement)~\cite{Battaglia:1999re}.
\item Select a sample of unbiased $\Ho\Z$ events (events in the recoil mass
      peak) and determine the fraction of events corresponding to a given
      $\Ho\to X$ decay within this sample.
\end{enumerate}

The latter method was first applied to Higgs branching ratio studies
in~\cite{LC-PHSM-2002-003}. Since in this approach binomial (or in principle
multi-nomial) statistics can be applied, smaller errors of the branching
ratios can be inferred for the same number of events than from a rate measurement. 
Although 
only relying on events with $\Z\to\ellell$, the latter method yields errors
very similar to those of the TDR method~\cite{Battaglia:1999re}. 
The achievable precision for
the both methods for a SM Higgs boson of 120 GeV from a sample of 500 $\fb$
is shown in Table \ref{tab:higgsbr}. A possible combination of both methods
is currently being investigated. While for the hadronic Higgs decays, there
is a sizable overlap, for the $\Ho\to W^+W^-$ decay a significant improvement
may be expected from combination.

\begin{figure}[htb]
\centering
\epsfig{height=\linewidth,file=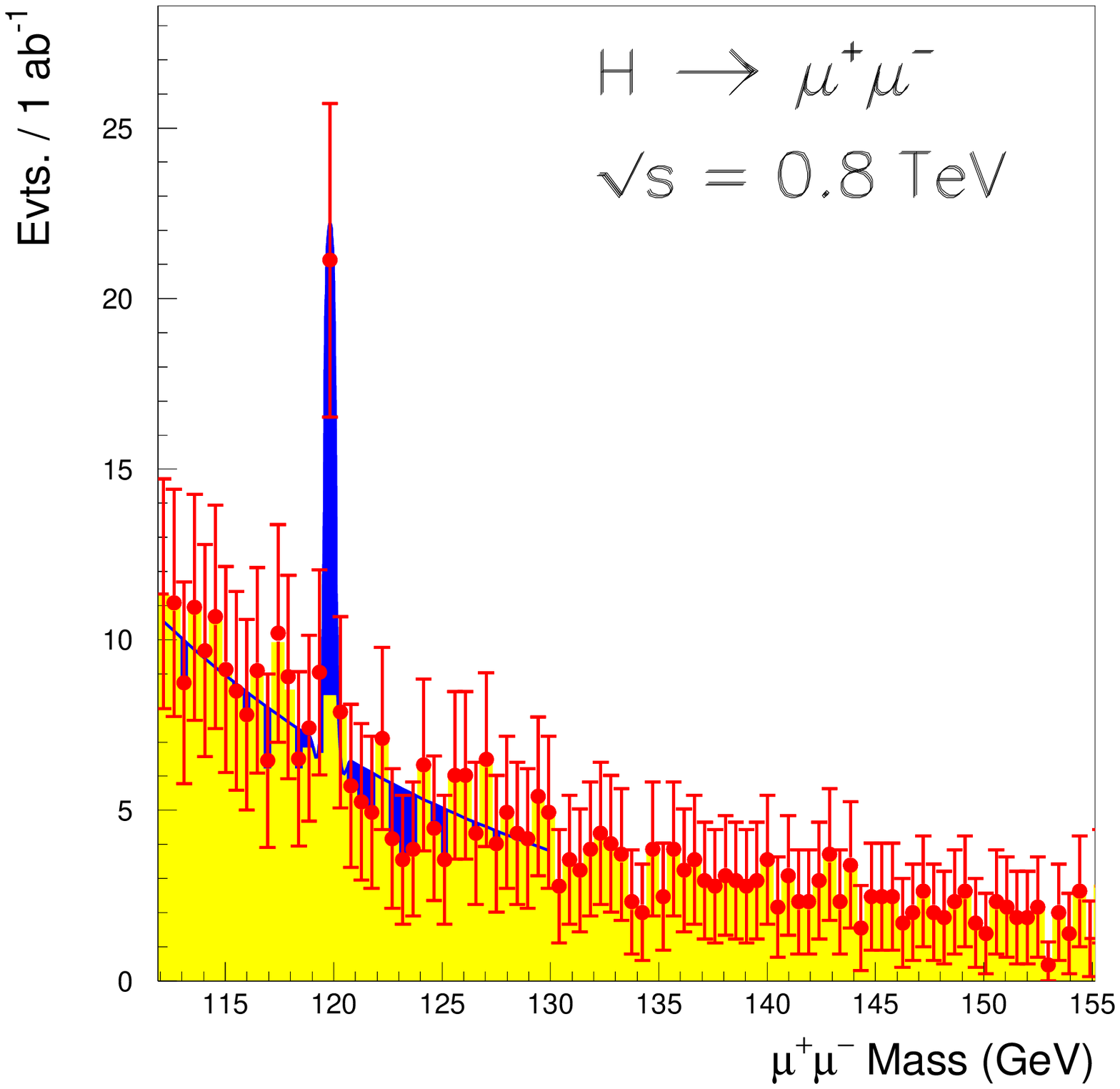}
\caption{Expected mass spectrum for the decay $\Ho\to\mu^+\mu^-$ from a sample of
1ab$^{-1}$ at $\sqrts = 800 $ GeV for $\mH = 120$ GeV (from~\cite{Battaglia:2001vf}).}
\label{fig:htomu}
\end{figure}

Besides the decays into 
$\bb, \cc, \glgl, \tautau, W^+W^-$, $Z^0 Z^0, $ and $\gamma\gamma$ further decay
modes have been studied. The very rare decay $\Ho\to\mu^+\mu^-$ might
be detectable in WW-fusion events at $\sqrt{s} = 800 $ GeV for $\mH = 120 $ GeV.
A measurement of the muon Yukawa coupling with
approximately 15\% relative accuracy may be obtained from
a sample of 1 ab$^{-1}$. Here,
the logarithmic rise of the signal cross-section with $\sqrt{s}$ is
of advantage. A precision measurement of the $\Ho\to\mu^+\mu^-$ branching
ratio however can only be performed at even higher luminosity or 
at higher energy~\cite{Battaglia:2001vf}.
The expected signal is shown in Fig.~\ref{fig:htomu}.

\begin{table*}[hbt]
\begin{center}
\caption{Summary of expected precisions on Higgs boson branching ratios 
from existing studies within the ECFA/DESY workshops. (a) for 500 fb$^{-1}$
at 350 GeV; (b) for 500 fb$^{-1}$ at 500 GeV; (c) for 1 ab$^{-1}$ at 500 GeV;
(d) for 1 ab$^{-1}$ at 800 GeV; (e) as for (a), but method described 
in~\cite{LC-PHSM-2002-003} (see text).}
\vspace{0.3cm}
\begin{tabular}{|l|r|r|r|r|r|r|r|r|r|}
\hline
Mass(GeV) & 120 & 140 & 160 & 180 & 200 & 220 & 240 & 280 & 320 \\ \hline
Decay     & \multicolumn{9}{c|}{Relative Precision (\%)} \\ \hline
$\bb$     & 2.4 (a) / 1.9 (e)& 2.6 (a)& 6.5 (a)&12.0 (d)& 17.0 (d)&28.0 (d)&     &     &     \\
$\cc$     & 8.3 (a) / 8.1 (e)&19.0 (a)&     &     &     &     &     &     &     \\
$\tau\tau$& 5.0 (a) / 7.1 (e)& 8.0 (a)&     &     &     &     &     &     &     \\ \hline
$\mu\mu$  &30. (d) &     &     &     &     &     &     &     &     \\ \hline
gg        & 5.5 (a) /4.8 (e)&14.0 (a)&     &     &     &     &     &     &     \\
WW        & 5.1 (a) / 3.6 (e)& 2.5 (a)& 2.1 (a)&     & 3.5 (b)&     & 5.0 (b)& 7.7 (b)& 8.6 (b)\\
ZZ        &     &     &16.9 (a)&     & 9.9 (b)&     &10.8 (b)&16.2 (b)& 17.3 (b)\\
$\gaga$   &23.0 (b) / 35.0 (e)&     &     &     &     &     &     &     &     \\
$\Z\gamma$&     &27.0 (c)&     &     &     &     &     &     &     \\ \hline
\end{tabular}
\label{tab:higgsbr}
\end{center}
\end{table*}

\begin{figure}[htb]
\centering
\epsfig{width=\linewidth,file=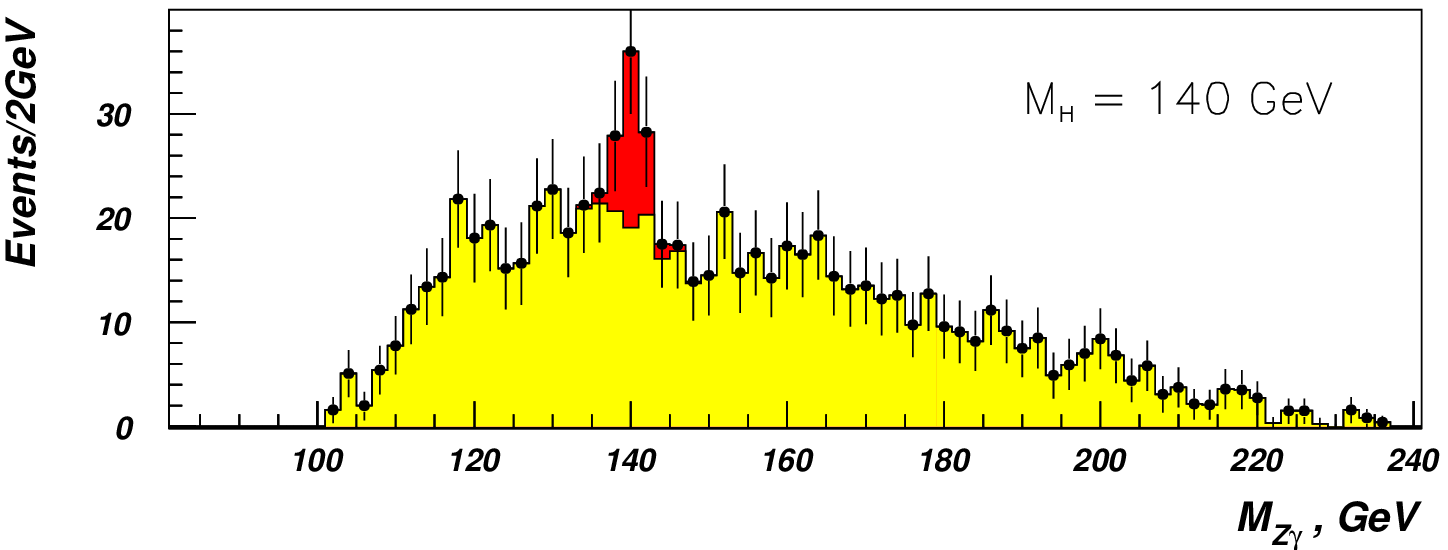}
\caption{Expected mass spectrum for the decay $\Ho\to\Z\gamma$ from a 
sample of 1 ab$^{-1}$ at $\sqrts = 500 $ GeV for $\mH = 120$ GeV
(from~\cite{LC-PHSM-2003-004}).}
\label{fig:zgamma}
\end{figure}

Another rare Higgs boson decay is the loop-induced $\Ho\to\Z\gamma$ decay.
This decay has been studied in the $WW\to\Ho\to\qq\gamma$ final state for
a sample of 1$\ab$ at 500~GeV for 120~GeV $< \mH < $ 160~GeV. Around the
expected maximum of the branching ratio for a SM Higgs boson (140~GeV),
a relative error of 27\% can be expected while for lower (120 GeV) and 
higher (160 GeV) Higgs masses 
only upper limits at 70-80\% of the SM branching ratio
can be expected to be set~\cite{LC-PHSM-2003-004}.
The expected signal is shown in Fig.~\ref{fig:zgamma} together with the background.

\subsection{Invisible Higgs Decays}

\begin{figure}[htb]
\centerline
{\epsfig{file=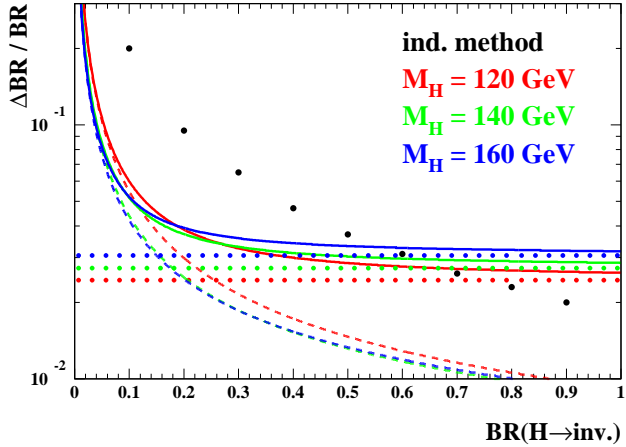,width=\linewidth}}
\caption{Accuracy on the branching ratio $\Ho\to invisible$, as a function of 
$BR(\Ho\to invisible$) for three Higgs masses using 500 $\fb$ at
350 GeV (full line). The dashed and dotted lines indicate the contributions
from the measurement of the invisible rate and from the total Higgs-strahlung
cross section measurement, respectively. The large dots are the result
of the indirect method, presented in the TDR (from ~\cite{schumacher}).}
\label{fig:invisible}
\end{figure}

In the TDR it was pointed out that the decay independent recoil mass
technique allows us to extract a possible invisible decay width of the
Higgs boson by comparing the rate of events in the recoil mass peak
with the rate for all visible decays. This indirect technique is now 
complemented by a study which explicitly asks for missing energy and
momentum compatible with an invisible Higgs decay. At $\sqrt{s} = 350 $~GeV,
the achievable precision on the invisible branching ratio is shown to be
significantly higher than in the indirect approach, yielding e.g.~a 
relative precision of $\sim 10 $\% for a branching ratio of 5\% 
and a 5$\sigma$ observation down to a branching ratio of 1.5-2.0\%
with 500 fb$^{-1}$ at $\sqrt{s} = 350 $~GeV and Higgs masses between 120 and
160 GeV~\cite{schumacher} (see Fig.~\ref{fig:invisible}).

\subsection{Heavier SM Higgs Boson}

Above a Higgs mass of approximately 2 $\mW$, the phenomenology of the
SM Higgs changes quite drastically. First, the bosonic decays into $\WW$ and $\Z\Z$
rapidly become dominant, leaving only very little room for Yukawa couplings
to be probed directly. Second, the total decay width increases rapidly 
with mass, exceeding 1 GeV for $\mH > 190 $~GeV.

In order to assess the question up to which Higgs mass a direct Yukawa
coupling measurement would still be possible, a study was performed which
aims at selecting $\Ho\to\bb$ as a rare Higgs decay~\cite{Battaglia-StMalo}.
Like in the case of $\Ho\to\mu^+\mu^-$, the large number of Higgs
bosons produced in the WW-fusion channel at high energy is favorable
in comparison to using the Higgs-strahlung process at lower energies.
For 1$\ab$ of data at $\sqrt{s} = 800 $ GeV, a 5$\sigma$ sensitivity
to the bottom Yukawa coupling is achievable for $\mH < 210 $ GeV.
A measurement of the branching ratio $BR(\Ho\to\bb)$ is possible with
(12,17,28) \% accuracy for $\mH = $ (180,200,220)~GeV.

\begin{figure}[htb]
\centering
\epsfig{width=\linewidth,file=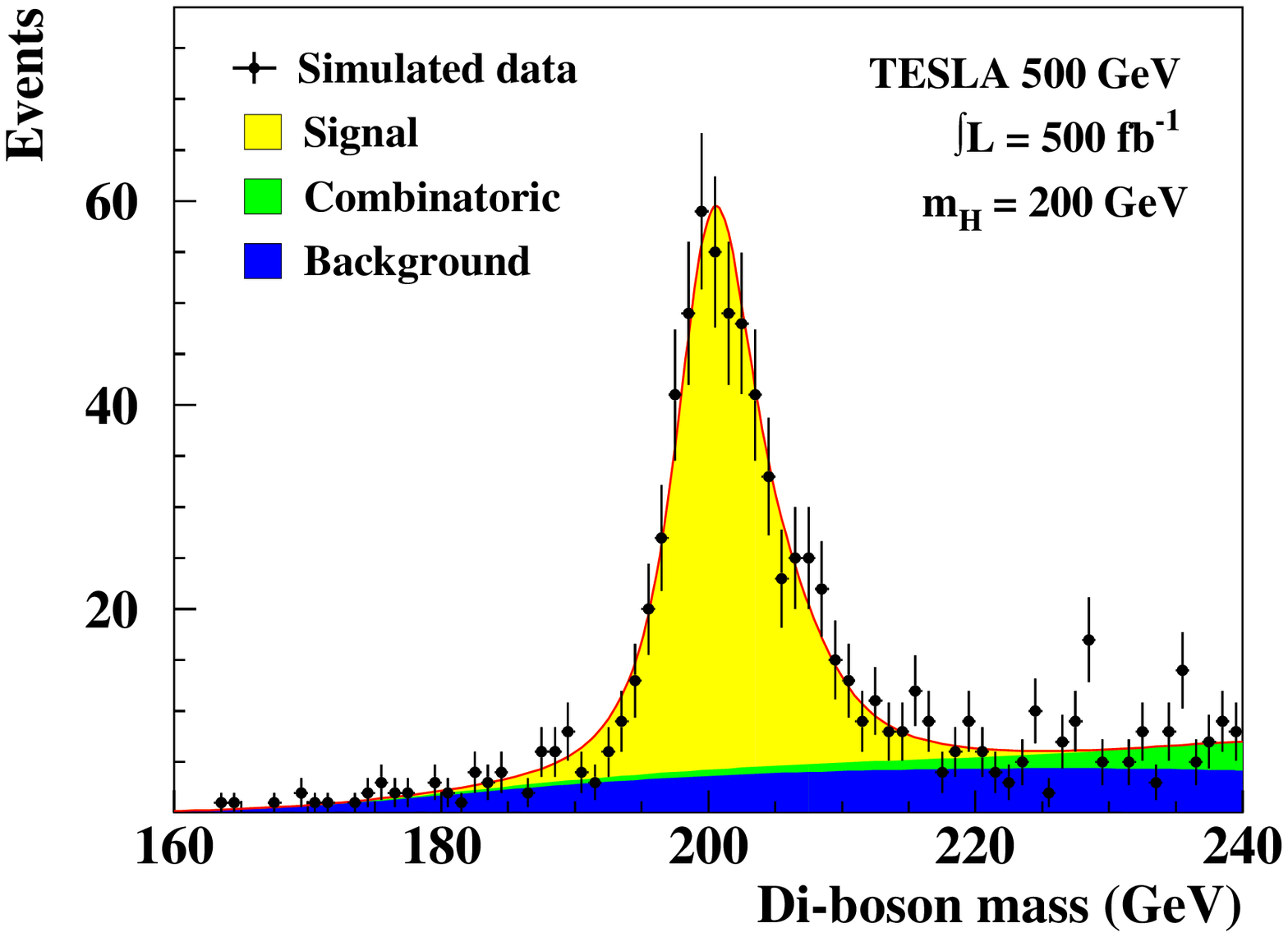}\\
\epsfig{width=\linewidth,file=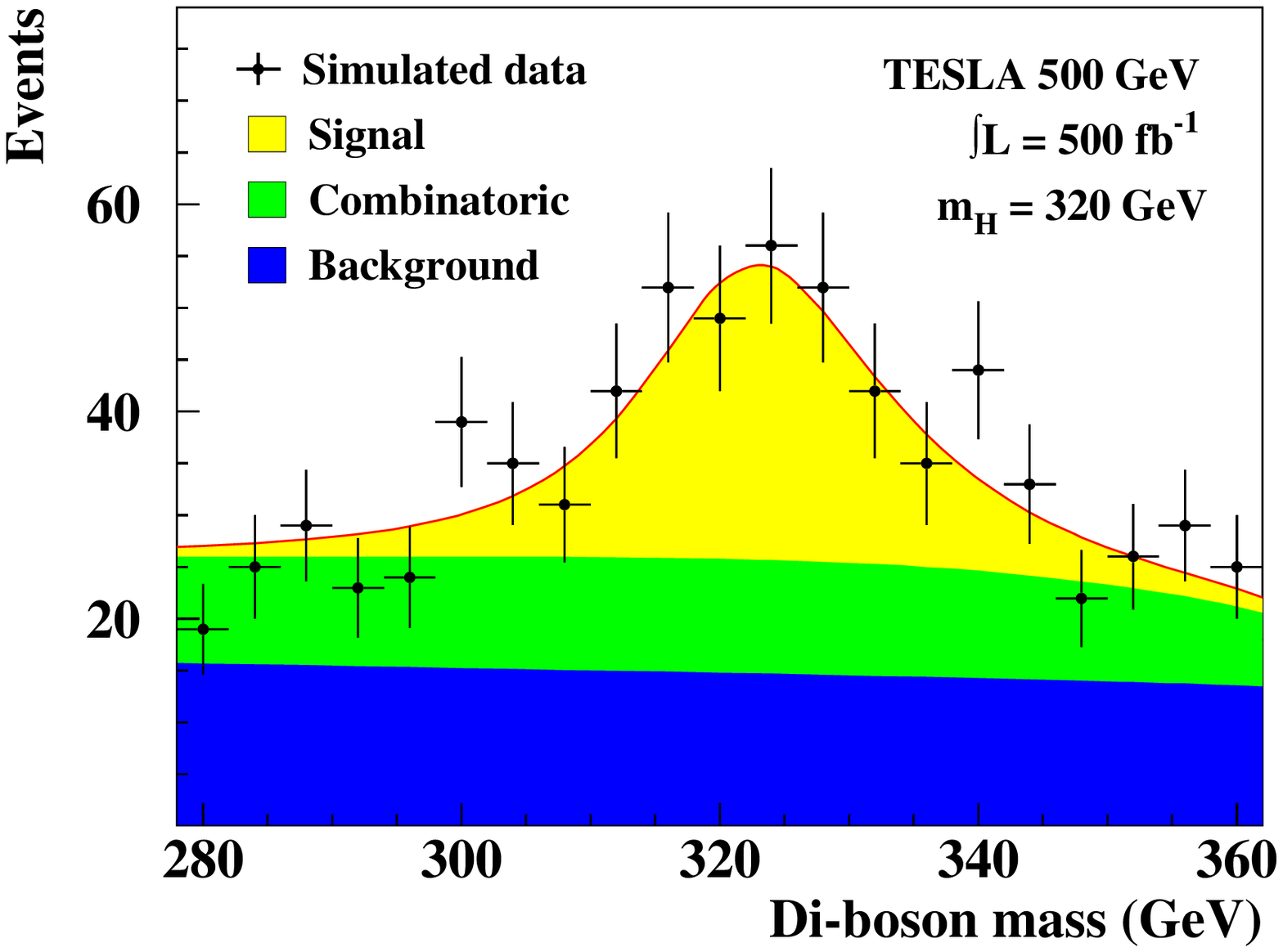}
\caption{Expected reconstructed Higgs boson mass spectra for
$\mH = 200 $ GeV and $\mH = 320 $ GeV from  500 $\fb$ at
500~GeV (from~\cite{LC-PHSM-2003-066}). }
\label{fig:niels}
\end{figure}

The second question about heavier Higgs bosons is, whether the Higgs
line-shape parameters (mass, decay width, Higgs-strahlung production cross
section) can be measured. A complete study of the mass range
200 GeV $< \mH <$320 GeV has been performed~\cite{LC-PHSM-2003-066}.
The final state $\qq\qq\ellell$ resulting from $\Ho\Z\to\Z\Z\Z$ and
from $\Ho\Z\to\WW\Z$ is selected. A kinematic fit is used to assign
the possible di-jet combinations to bosons ($\WW$ or $\Z\Z$). The resulting
di-boson mass spectrum can be fitted by a Breit-Wigner distribution
convoluted with a detector resolution function. A relative uncertainty
on the Higgs mass of 0.11 -- 0.36 \% is achievable from 500 $\fb$ at
500~GeV for masses between 200 and 320~GeV. The resolution on the total
width varies between 22 and 34\% for the same mass range. Finally, 
the total Higgs-strahlung cross-section can be measured with 3.5 -- 6.3\%
precision. Under the assumption that only $\Ho\to\WW$ and $\Ho\to\Z\Z$
decays are relevant, their branching ratios can be extracted with
3.5--8.6\% and 9.9--17.3\%, respectively (see Table~\ref{tab:heavyh}).
The expected mass spectra for $\mH = 200 $ GeV and $\mH = 320 $ GeV are
shown in Fig.~\ref{fig:niels}. 

\begin{table}[hbt]
\begin{center}
\caption{Expected precision on Higgs boson line-shape parameters 
for $200 < \mH < 320$ GeV at a LC with $\sqrt{s} = 500$ GeV. }
\vspace{0.3cm}
\begin{tabular}{|c|c|c|c|}
\hline
$\mH$ (GeV) & $\Delta\sigma$ (\%) & $\Delta\mH$ (\%) & 
                                                  $\Delta\Gamma_{\mathrm{H}}$ (\%) 
                                                  \\ \hline
200 & 3.6 & 0.11& 34   \\ 
240 & 3.8 & 0.17& 27   \\
280 & 4.4 & 0.24& 23   \\
320 & 6.3 & 0.36& 26   \\ \hline

\end{tabular}
\label{tab:heavyh}
\end{center}
\end{table}

\subsection{Top Yukawa Coupling}

\begin{figure*}[htb]
\centering
\epsfig{width=0.9\linewidth,file=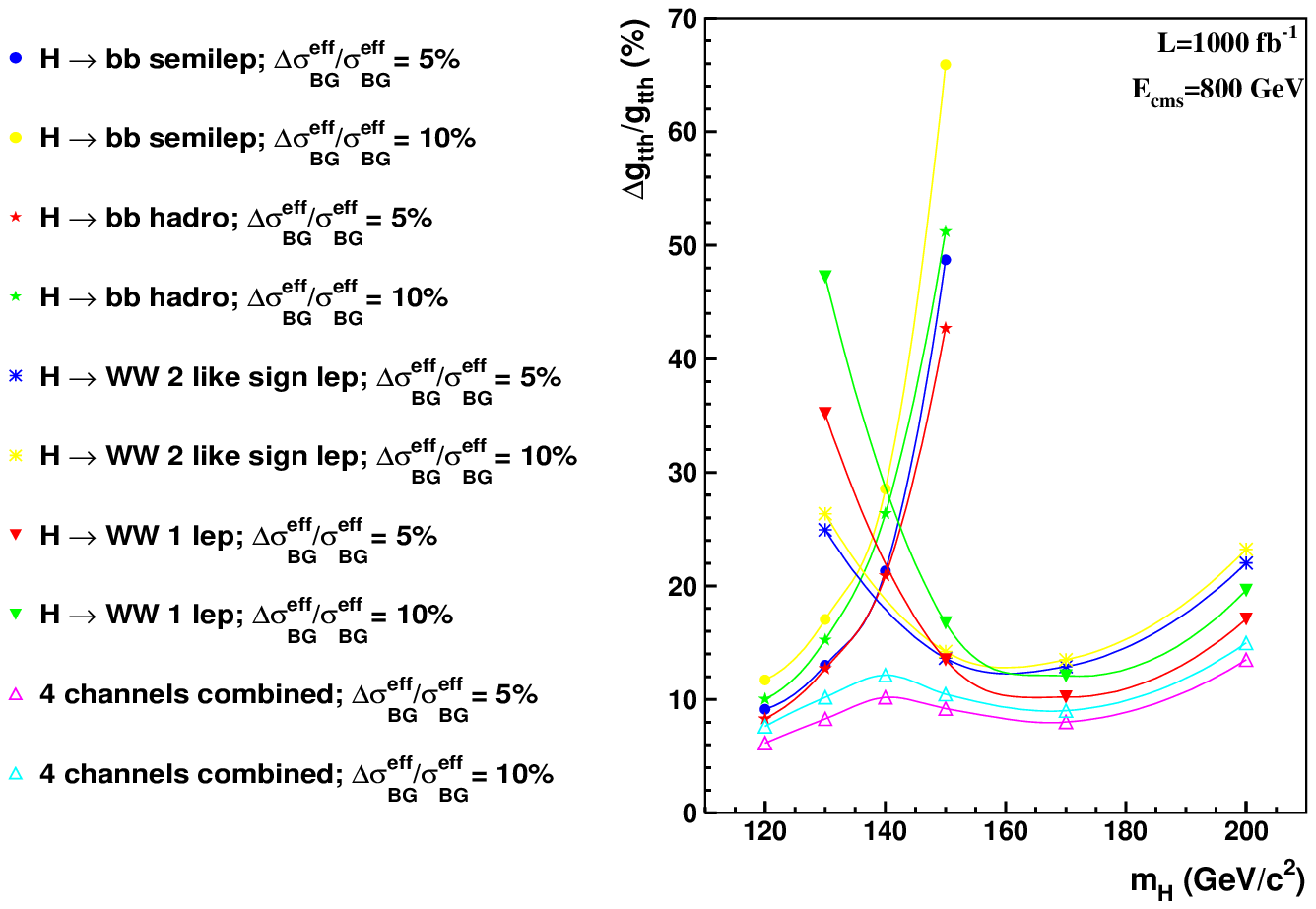}
\caption{Expected relative precision on the top Yukawa coupling
for $120 < \mH < 200 $ GeV from  1 $\ab$ at
800~GeV for various final states and for two different assumptions of
the systematic uncertainty on the background (BG) normalization (from~\cite{gaynew}).}
\label{fig:tth}
\end{figure*}

For $\mH < 2 \mt$, the top quark Yukawa coupling is not directly accessible from
Higgs decays. The only relevant tree level process to access the top
quark Yukawa coupling
is the process $\ee\to\Ho\ttbar$~\cite{tthtree}. 
Due to the large masses
of the final state particles, the process only has a significant cross-section
at center-of-mass energies significantly beyond 500~GeV. 
Higher order QCD corrections to the process have been calculated and are
significant~\cite{tthqcd}. Recently, also the full ${\cal O} (\alpha)$
electro-weak corrections became available~\cite{tthew}.
Experimental studies have been performed for $\mH < 130 $ GeV 
in the TDR~\cite{juste} and in the NLC study~\cite{tthbaer}.
Recently a completely new study has been performed with refined b-tagging
simulation as well as for an extended mass range of up to $\mH = 200 $ GeV,
exploiting also the $\Ho\to\WW$ decay~\cite{gaynew}.
For the $\Ho\to\bb$ case, both the $\ttbar\ra\bb\qq\lnu$ and the 
$\ttbar\ra\bb\qq\qq$ channels have been analyzed. For the $\Ho\to\WW$ case,
the 2-like-sign lepton plus 6-jet 
and the single lepton plus 8-jet final states were studied. The events
were selected by neural networks. The generic 6-fermion background is
fully taken into account.
The expected uncertainties on the
top Yukawa coupling for 1$\ab$ at 800 GeV range from 
6--14\% for $120 < \mH < 200$ GeV and are shown in Fig.~\ref{fig:tth}.

\subsection{Higgs Potential}

\begin{figure}[htb]
\centering
\epsfig{height=0.95\linewidth,file=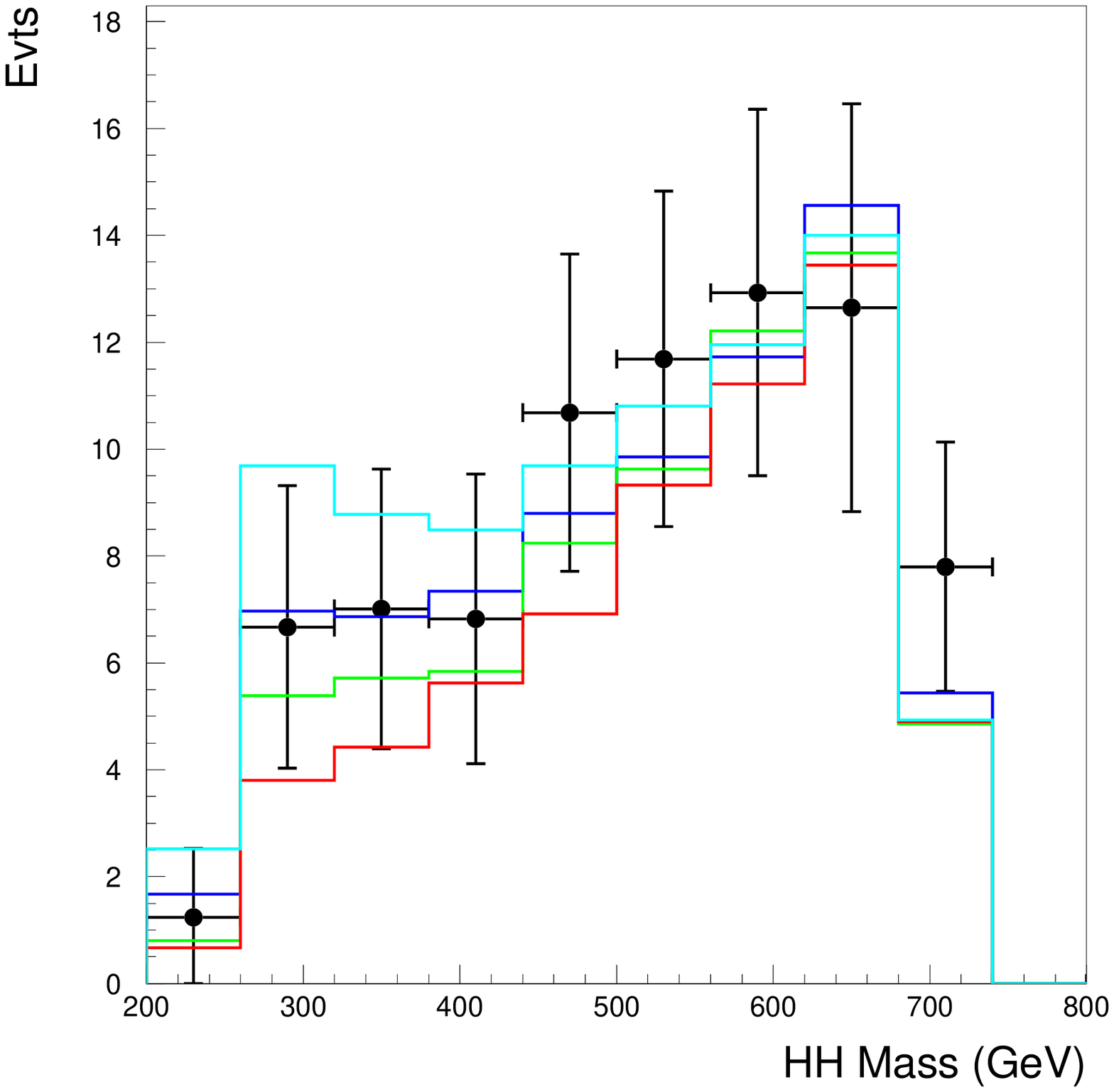}
\caption{Distribution of the $\Ho\Ho$ invariant mass in $\ee\to\Ho\Ho\Z$
events for $\mH = 120 $ GeV (1 $\ab$ at
800~GeV). The histograms are for predictions of the trilinear Higgs coupling
ranging from 1.25 to 0.5 (top to bottom) times the SM coupling. (from~\cite{Battaglia:2001nn}). }
\label{fig:potential}
\end{figure}

The observation of a non-zero self-coupling of the Higgs boson is the
ultimate proof of spontaneous symmetry breaking being responsible for
mass generation of the SM bosons and fermions since it probes the
shape of the Higgs potential and thus the presence of a vacuum expectation
value. Higgs boson self-coupling in general leads to triple and quartic
Higgs boson couplings out of which only the former is accessible. For
500 GeV center-of-mass energy, the double Higgs-strahlung process,
$\ee\to\Ho\Ho\Z$ is most promising for observation, the small cross-section
of 0.1~-~0.2~fb however demands the highest possible luminosity and 
calls for ultimate jet energy resolution since only if the most frequent
six jet final state $\bb\bb\qq$ can be reconstructed, the signal rate becomes
significant. The cross-section has been calculated in~\cite{maggie} and
radiative corrections became known recently~\cite{hhzew}.
In the TDR, an experimental analysis for $\mH = 120 $ GeV was 
presented~\cite{gaylutz}
which concluded that with 1$\ab$ of data at 500~GeV, a precision of 
17 - 23 \% for $120 < \mH < 140 $ GeV on the $\ee\to\Ho\Ho\Z$ cross-section
can be achieved. Recently, the potential
of the WW-fusion channel for higher Higgs boson masses at higher energies
was discussed and compared to the possibilities at the LHC in~\cite{rainwater}.
Furthermore, it was discussed how the existing analyses might be 
improved by exploiting kinematic differences between the signal diagram
and diagrams which lead to the same final state without involving the
triple Higgs coupling (dilution diagrams), namely the sequential radiation
of two Higgs bosons from one Z boson and the diagram which involves the
quartic ZZHH coupling~\cite{Battaglia:2001nn}. 
In particular, the invariant mass of the hadronic system which is formed
by the two Higgs boson decay products is sensitive to the different
contributions to the HHZ final state. Its distribution is shown
in Fig.~\ref{fig:potential}. A reduction of the uncertainty on the
trilinear coupling from 0.23 to 0.20 can be obtained.

\section{MINIMAL SUPERSYMMETRIC HIGGS SECTOR}

\subsection{Theoretical Predictions}

\begin{figure}[htb]
\centering
\epsfig{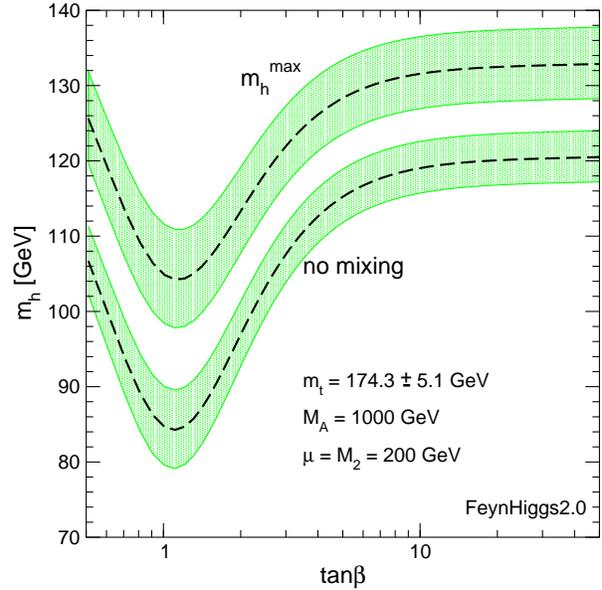}
\caption{Largest mass of the light CP even Higgs boson of the 
MSSM as a function of $\tb $ for two scenarios of scalar-top mixing
(no-mixing and $\mlh^{\mbox{max}} $ 
scenarios of~\cite{Carena:2002qg}). The bands indicate
the effect of varying the top quark mass by 1 standard deviation of its current error.}
\label{fig:mhmax}
\end{figure}

The Higgs sector of the Minimal Supersymmetric Standard Model (MSSM)
comprises two complex scalar field doublets which acquire vacuum expectation
values $v_1$ and $v_2$. After electro-weak symmetry breaking, 
two charged
Higgs bosons ($\Hpm$) and three neutral Higgs bosons emerge, two of which are
CP-even ($\lh, \hh$) and one is CP-odd ($\Ao$), if CP is conserved.
In contrast to the SM, the Higgs masses are predicted in terms of the
fundamental parameters of the MSSM. At tree level, the mass spectrum is
determined by $\tan\beta = v_2 / v_1$ and $\mA$ and the mass
of the $\lh$ has to fulfill $\mlh < \mZ$. Higher order corrections, 
predominantly from loops involving third generation fermions and their
supersymmetric partners, have large influence. In particular, $\mlh$ 
can be as large as 135 GeV~\cite{mh135}. A compilation of more recent higher order
corrections can be found in~\cite{mssmradcor}.
The value of $\mlh$ as a function of $\tb$ is shown for two different
cases of scalar-top mixing (no-mixing and $\mlh^{\mbox{max}} $
scenarios of~\cite{Carena:2002qg})
in Fig.~\ref{fig:mhmax}.
The complete 1-loop and dominant 2-loop SUSY corrections 
to the production cross-sections for
$\ee\to\lh\Z$~\cite{susyhz,susyhz2} and the 1-loop corrections from stop-sbottom
loops for
$\ee\to\nu_e\bar{\nu_e}\lh$~\cite{LC-TH-2002-019, eberl}
are calculated.

The MSSM Higgs sector exhibits a so-called decoupling limit as $\mA$ becomes
large, in which the $\lh$ approaches the properties of the 
SM Higgs boson~\cite{haberdecoupling}
This limit is approached relatively fast for $\mA > 200 $ GeV in a large portion
of the MSSM parameter space. However, also scenarios far away from 
decoupling (e.g.~{\it the intense coupling
scenario}~\cite{intense}) is experimentally not excluded and theoretically
possible. In such a scenario, all Higgs bosons are accessible already
at 500 GeV and a rich phenomenology is waiting to be disentangled.
The closer the MSSM scenario moves towards the decoupling limit the more
difficult it becomes to distinguish the Higgs sector from the SM. 
Therefore most analyses
focus on a close-to-decoupling scenario. In this case, the analyses for
a light SM Higgs apply also for $\lh$. It is the task of the LC to
employ the precise measurements of the properties of this lightest Higgs
boson to distinguish it from a SM Higgs and draw conclusions on the 
supersymmetric parameters.

\subsection{Study of Heavy Neutral SUSY Higgs Bosons}

\begin{figure}[htb]
\centering
\epsfig{height=0.90\linewidth,file=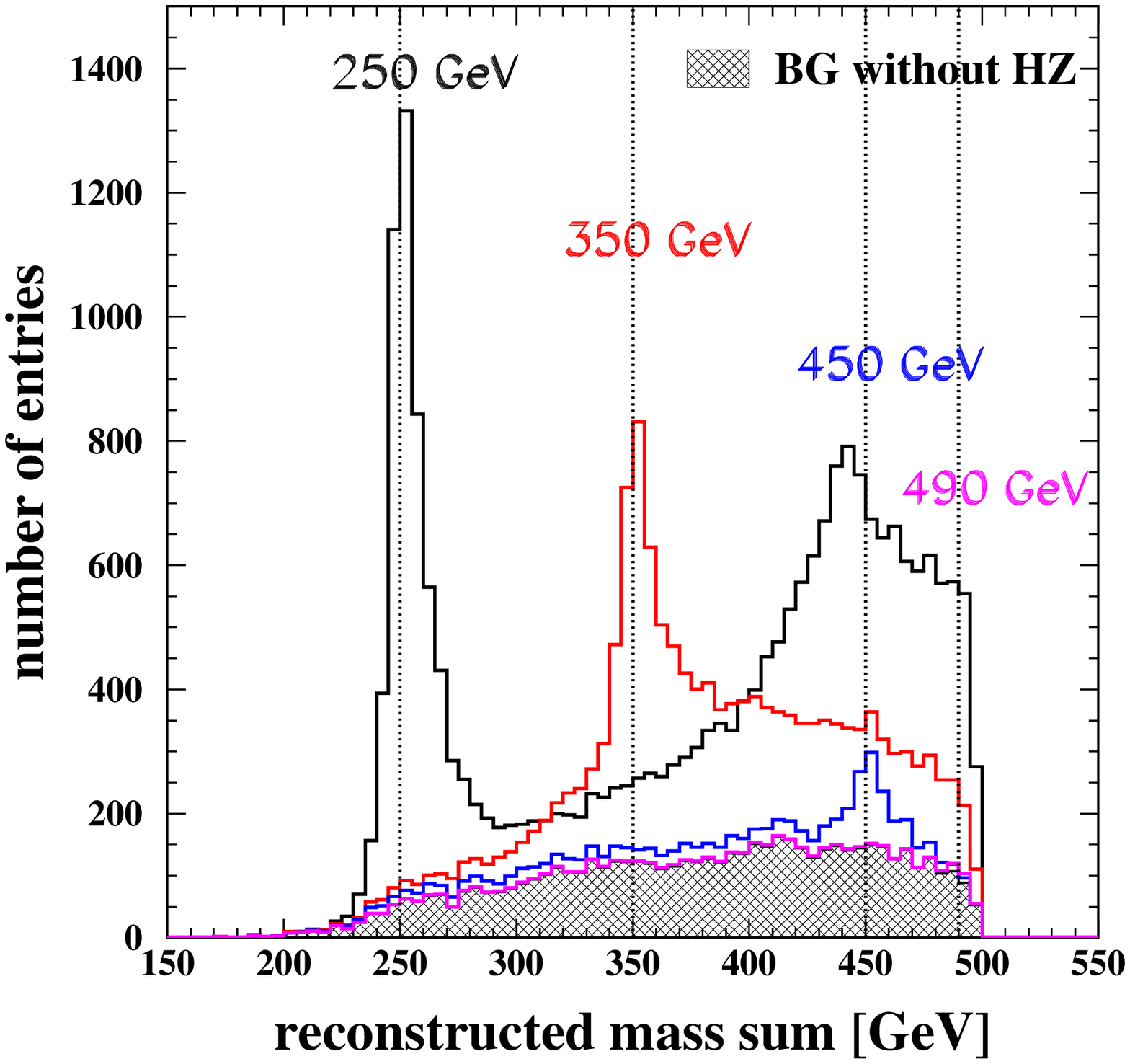} \\
\epsfig{height=0.90\linewidth,file=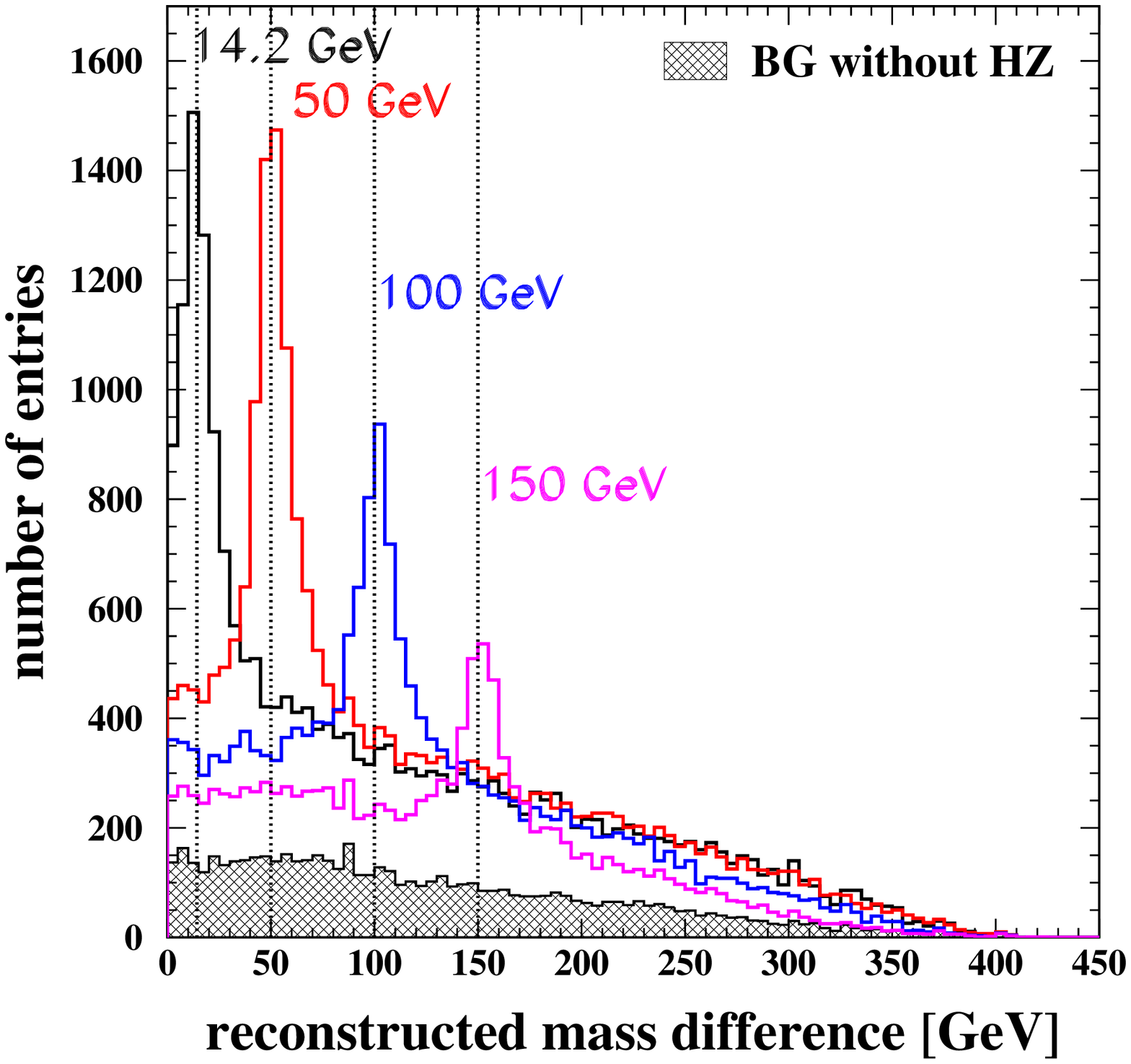}
\caption{Simulated signals and background of the process $\ee\ra\Ho\Ao\to
\bb\bb$. Top: reconstructed sum of the two Higgs candidate masses.
Bottom: reconstructed difference of the two Higgs boson candidate masses.
The study was performed at 500 GeV center-of-mass energy and for 500$\fb$.
BR($\Ho\to\bb$) $=$ BR($\Ao\to\bb$) $= 0.9$ was assumed (from~\cite{klimk}).}
\label{fig:habbbb}
\end{figure}

\begin{figure}[htb]
\centering
\epsfig{height=0.90\linewidth,file=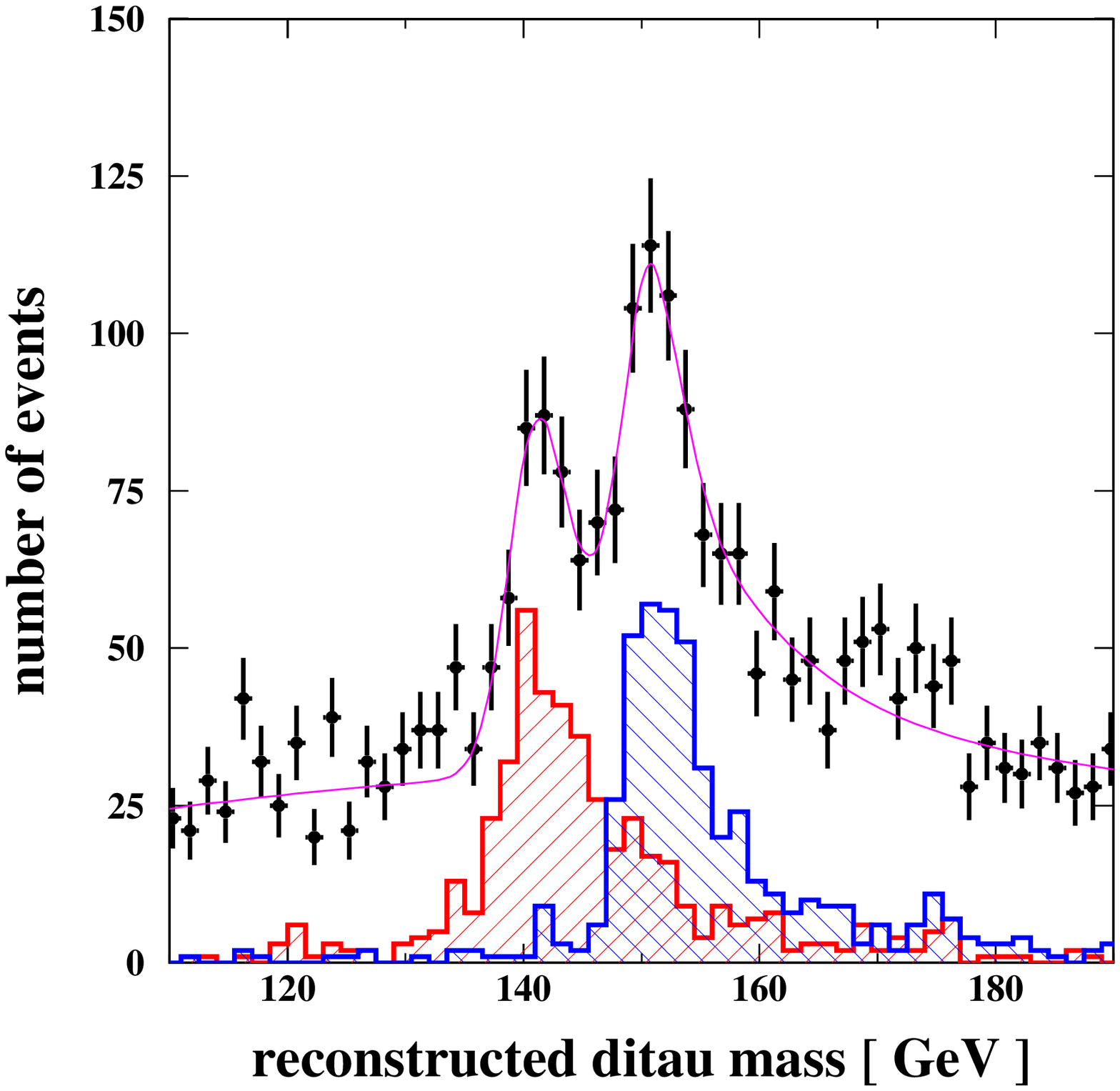} \\
\epsfig{height=0.90\linewidth,file=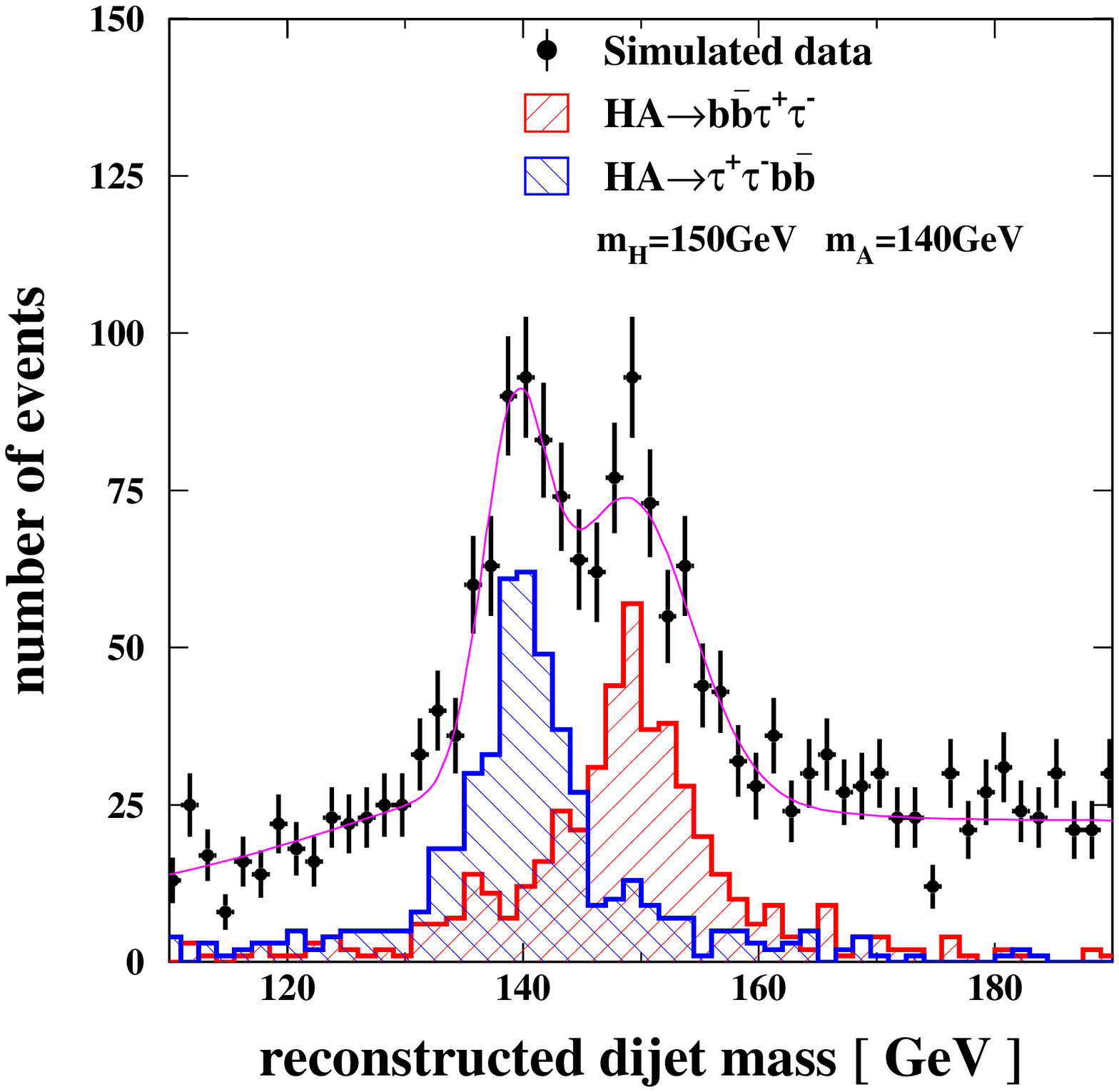}
\caption{Simulated signal and background of the process $\ee\ra\Ho\Ao\to
\bb\tautau (\tautau\bb)$ for $\mA = 140 $ GeV and $\mH = $ 150 GeV at
500 GeV center-of-mass energy (500$\fb$).
Top: reconstructed $\tau\tau$ invariant mass from a kinematic fit.
Bottom: reconstructed $\bb$ invariant mass from a kinematic fit.
BR($\Ho\to\tautau$) $=$ BR($\Ao\to\tautau$) $= 0.1$ was assumed 
(from~\cite{klimk}).}

\label{fig:habbtt}
\end{figure}

If $\cos^2{(\beta-\alpha)}$ is small\footnote{$\alpha$ is the mixing angle in the
CP-even neutral Higgs sector}, the heavy neutral MSSM Higgs bosons
are predominantly produced through the process, $\ee\ra\Ho\Ao$. With the mass
splitting between $\Ho$ and $\Ao$ being small for a large part of the 
parameter space, the mass reach of the LC for $\Ho$ and $\Ao$ is approximately
$\sqrt{s} / 2$. In this case, the coupling of the $\Ho$
to gauge bosons is small, therefore the dominant decays of both $\Ho$ and 
$\Ao$ are $\bb$ and $\tautau$. During the workshop, 
a new experimental study was started to fully determine the sensitivity
of the LC to the heavy MSSM Higgs bosons through the pair production 
process~\cite{klimk}.
For the first time, both the $\bb\bb$ and $\bb\tautau$ final states
are analyzed including detector simulation and complete standard
model backgrounds.
Preliminary results at 500 GeV and 800 GeV center-of-mass
energy were obtained.
The following assumptions are made: 500 $\fb$ at 500 GeV and at 800 GeV,
$\cos^2{(\beta-\alpha)} = 0$,
BR($\Ho\ra\bb$) = 90\%, BR($\Ho\ra\tautau$) = 10\%.
Mass reconstruction is performed using a kinematic fit which imposes
energy-momentum conservation. Therefore a good mass reconstruction is
achieved both in the $\bb\bb$ and $\bb\tautau$ final states, see 
Fig.~\ref{fig:habbbb} and~\ref{fig:habbtt}. The achievable precisions
on masses and topological cross-sections are listed in 
Table~\ref{tab:heavysusy} for various choices of $\mH$ and $\mA$.

\begin{table*}[hbt]
\begin{center}
\caption{Expected precision on the properties of heavy MSSM Higgs bosons
from 500 fb$^{-1}$ at 500 GeV (a) and 800 GeV (b), respectively 
(from~\cite{klimk}).}
\vspace{0.3cm}
\begin{tabular}{|c|c|c|c|c|c|c|}
\hline
& $\mA$ & $\mH $ & \multicolumn{4}{c|}{precision on} \\ \hline
&       &       & $(\mA+\mH )  $ & $ (|\mA - \mH |)$ & 
$\sigma(\bb\bb)$ & $\sigma(\bb\tautau)/\sigma(\tautau\bb)$ \\ \hline
    &(GeV)&(GeV)&(GeV)&(GeV)& (\%)& (\%) \\ \hline
(a) & 140 & 150 & 0.2 & 0.2 & 1.5 & 7.2/6.3 \\ \hline
(a) & 150 & 200 & 0.3 & 0.4 & 2.3 & 9.7/8.7 \\ \hline
(a) & 200 & 200 & 0.4 & 0.4 & 2.7 & 8.1 \\ \hline
(a) & 200 & 250 & 0.4 & 1.2 & 6.5 & - \\ \hline
(b) & 250 & 300 & 0.5 & 0.7 & 3.0 & 13.8/11.9 \\ \hline
(b) & 300 & 300 & 0.6 & 0.7 & 3.5 & 10.0 \\ \hline
(b) & 300 & 400 & 1.9 & 2.8 & 7.0 & - \\ \hline
\end{tabular}
\label{tab:heavysusy}
\end{center}
\end{table*}

\begin{figure}[htb]
\centering
\epsfig{width=\linewidth,file=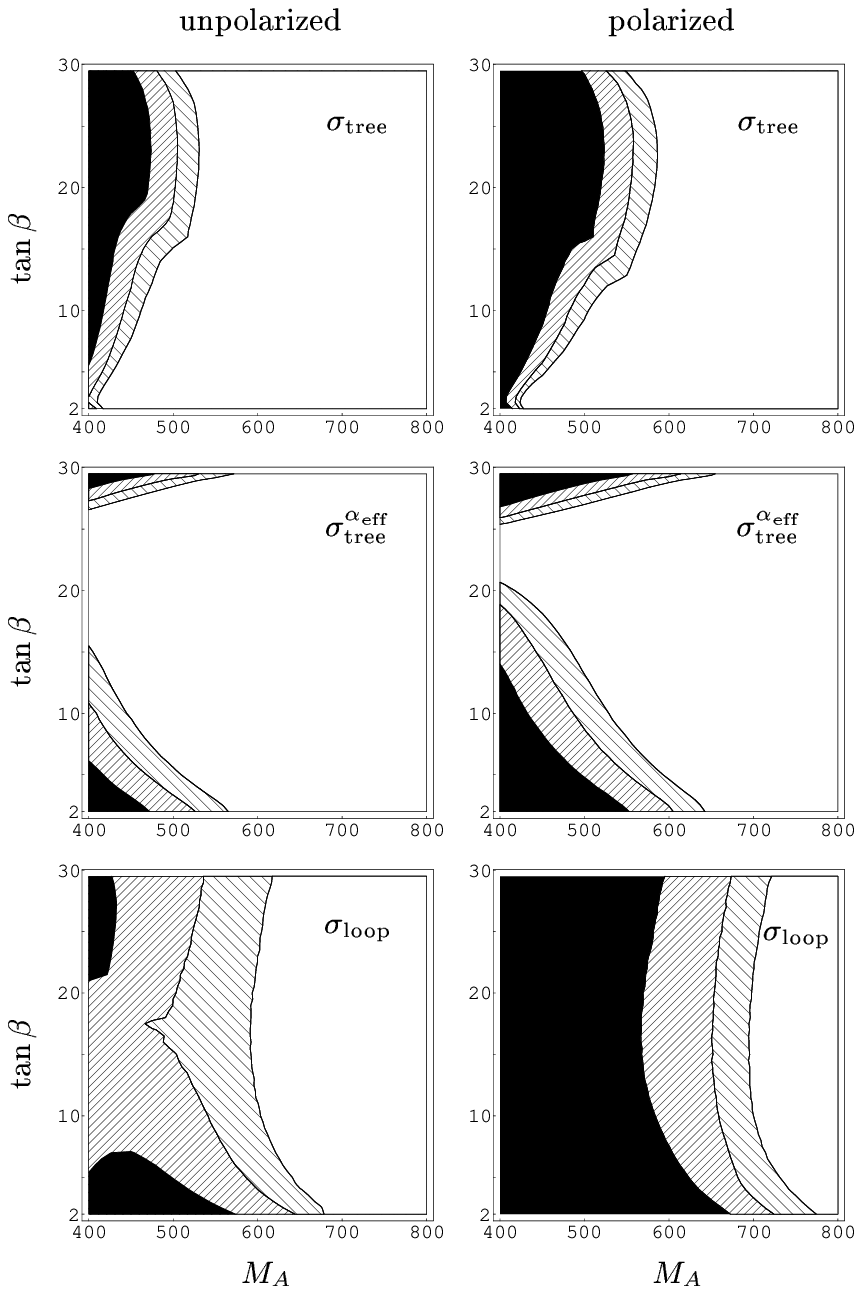,clip=}
\caption{Cross-section contours for $\ee\to\Ho \nu\bar{\nu}$
for a particular MSSM scenario (see text) 
in the $\mA-\tb$-plane for $\sqrts = 1 $ TeV.
The different shadings correspond to:
white: $\sigma \le 0.01 \fbc $, light shaded: $0.01 \fbc \le \sigma \le 0.02 \fbc$, 
dark shaded: $0.02 \fbc \le \sigma \le 0.05 \fbc$, black: $\sigma \ge 0.05 \fbc$
(from~\cite{LC-TH-2002-019}). The left figure is for unpolarized beams, the
right figure for an electron (positron) polarization of 0.8 (0.6).}
\label{fig:heavymssm}
\end{figure}

Since at the tree level and in the decoupling limit the heavy neutral MSSM
Higgs bosons decouple from the $\Z$, the mass reach for their discovery at
a LC is limited to approximately $\sqrts / 2$ from the pair production
process. It has been investigated during the workshop, how single production
mechanisms could extend the mass reach of an $\ee$ LC. In particular,
the WW-fusion process $\ee\to\nu_e\bar{\nu_e}\Ho$ 
has been investigated~\cite{LC-TH-2002-019}. 
Its tree level cross-section is proportional to
$\cos(\beta-\alpha)$. Depending
on the SUSY parameters, radiative corrections might increase the cross-section
for $\ee\to\nu_e\bar{\nu_e}\Ho$, possibly allowing discovery beyond the 
pair production kinematic limit for certain choices of the MSSM parameters.
Using left-polarized electron beams and right-polarized positron beams
the cross-section can further be enhanced.
A particular scenario where this is the case has been chosen 
in~~\cite{LC-TH-2002-019} ($M_{\mathrm{SUSY}} = $ 350 GeV, $\mu = 1000 $ GeV,
$M_2 = 200 $ GeV and large stop mixing). Cross-section contours for this
scenario are shown in Fig.~\ref{fig:heavymssm}. 

\subsection{Charged Higgs Bosons}

\begin{figure}[htb]
\centering
\epsfig{height=0.95\linewidth,file=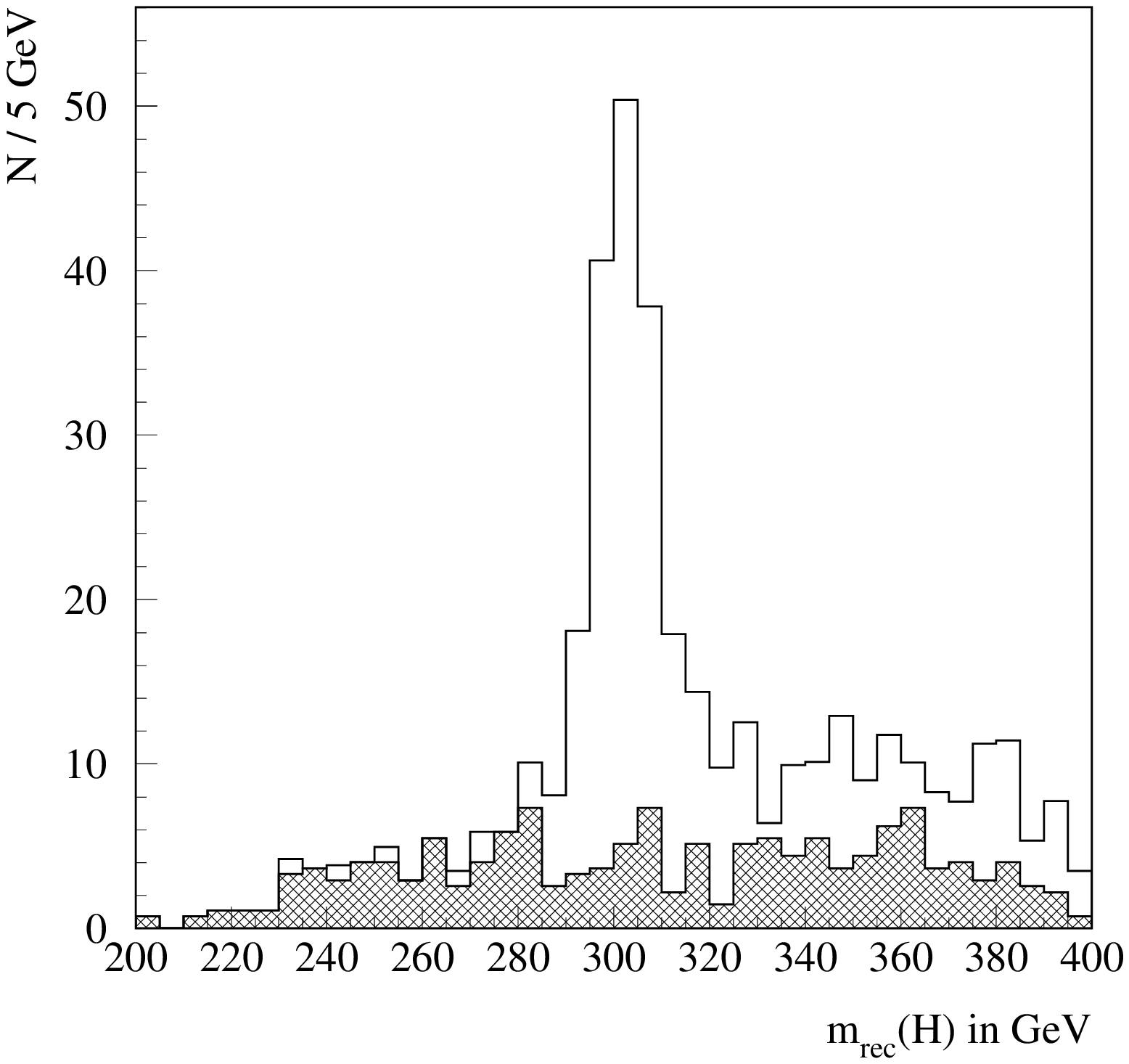}
\caption{Simulated signal and background of the process $\ee\ra\Hp\Hm\to
t\bar{b}\bar{t}b$ for $\mHpm = 300 $ GeV at
800 GeV center-of-mass energy (1$\ab$) (from~\cite{Battaglia:2001be}).}
\label{fig:hphm}
\end{figure}

Charged Higgs bosons can be pair-produced at the LC via $\ee\ra\Hp\Hm$ 
if $\mHpm < \sqrt{s} / 2$. A complete simulation of this process for
the decay $\Hp\ra t\bar{b}$ has been performed for $\sqrt{s} = 800 $ GeV,
1 ab$^{-1}$, and $\mHpm = 300 $ GeV~\cite{Battaglia:2001be}.
The expected signal and background are shown in Fig.~\ref{fig:hphm}. 
The mass resolution is approximately 1.5\%. A 5$\sigma$ discovery will
be possible for $\mHpm < 350 $ GeV.

\begin{figure}[htb]
\centering
\epsfig{height=0.94\linewidth,file=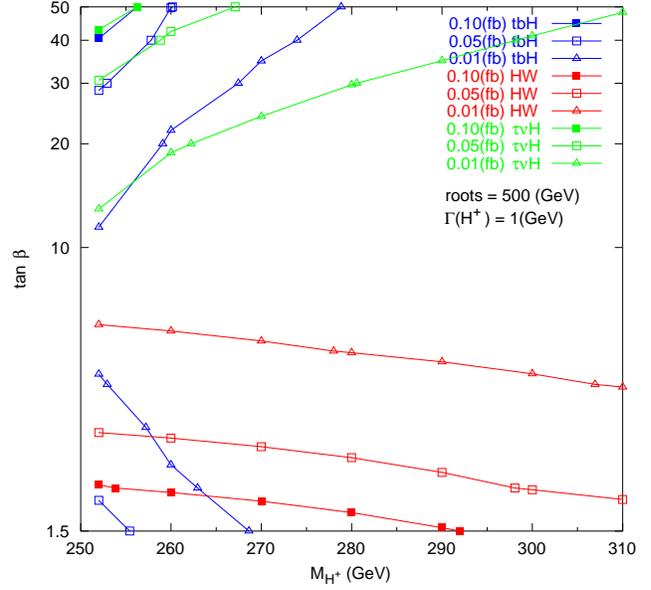} \\[0.5cm]
\epsfig{height=0.94\linewidth,file=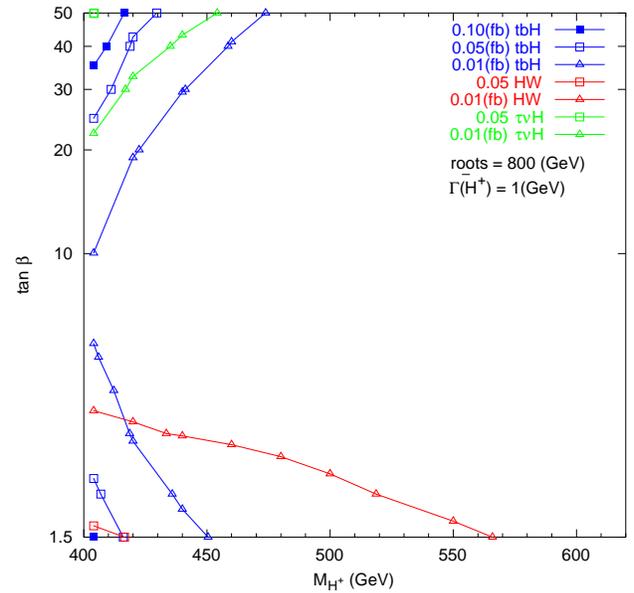}
\vspace{-0.2cm}
\caption{Cross-section contours for the processes
$\ee\to b\bar{t} H^+ (blue/dark), 
\ee\to \tau^- \bar{\nu_\tau} H^+$ (green/light grey),
and $\ee\to W^- H^+$ (red/medium grey) 
at $\sqrts = 500 $ GeV (upper) and at 800 GeV (lower).
(from~\cite{kiyoura}).}
\label{fig:singlecharged}
\end{figure}

Since in pair production the mass reach for charged Higgs bosons is 
limited to $\sqrts / 2$, also the rare processes of single charged Higgs
production may be considered. The dominant processes for single charged Higgs
production are $\ee\to b\bar{t} H^+, \ee\to \tau^- \bar{\nu_\tau} H^+$,
and $\ee\to W^- H^+$. Their cross-sections have been calculated at leading
order in~\cite{LC-TH-2002-010}. QCD corrections to $\ee\to b\bar{t} H^+$
have recently become available~\cite{kniehlhpm} and are sizable.
In general, parameter regions for which the production cross-section
exceeds 0.1~fb are rather small
for charged Higgs masses beyond the pair production threshold.
Cross-section contours for $\sqrts = 500 $ GeV and 800 GeV are shown in
Fig.~\ref{fig:singlecharged}.

\subsection{Constraints on SUSY Parameters}

\begin{figure}[htb]
\centering
\centerline{
\epsfig{width=\linewidth,file=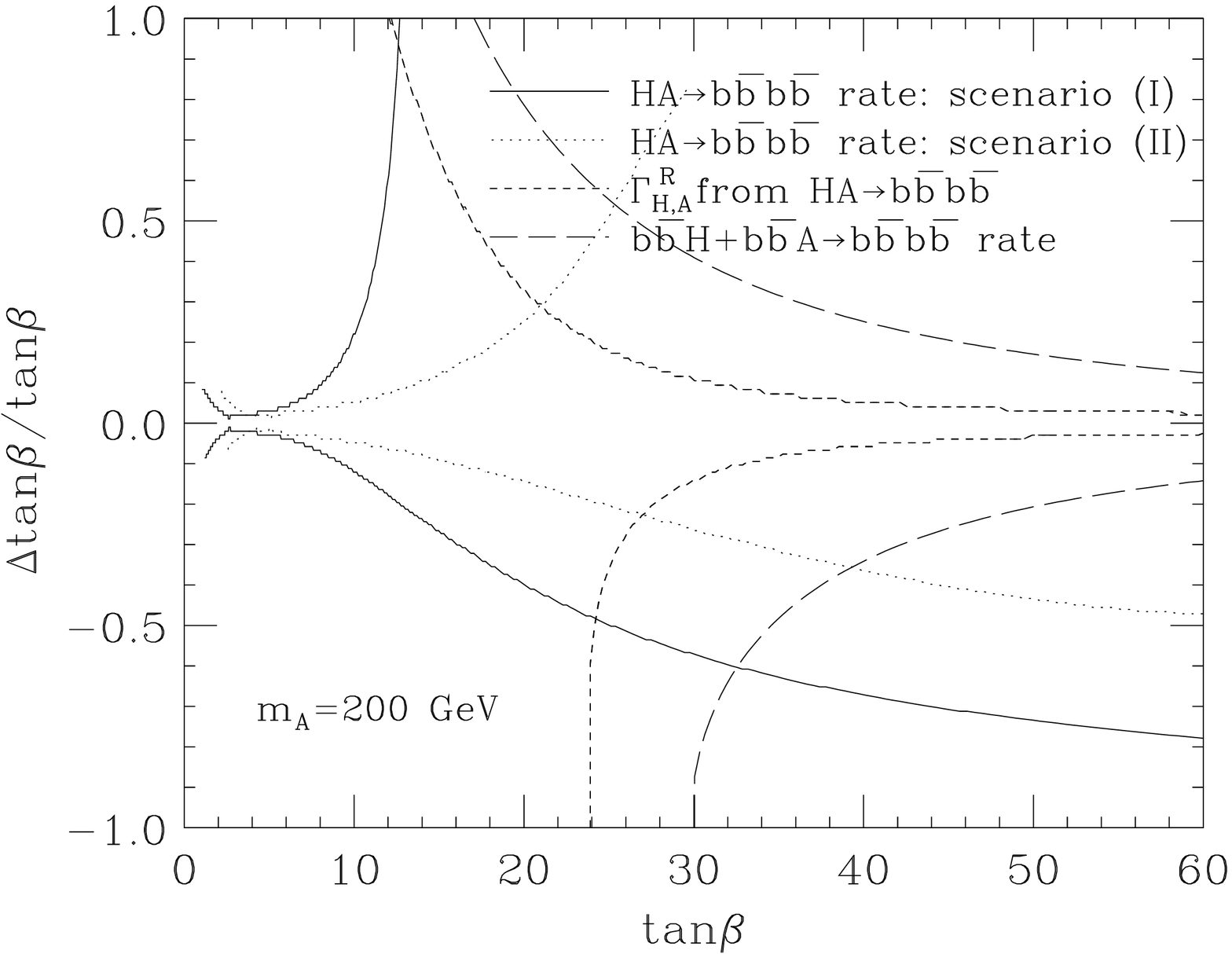}}
 \caption{
Sensitivity of the $ \ee\to\Ho\Ao\to\bb\bb $ and 
$\ee\to\bb\Ao/\Ho $ cross-sections and the total decay widths $\Gamma_{H/A} $
on $\tb$. Assumed measurement errors are for 2 $\ab$ at 500 GeV without
detector simulation (except for $\ee\to\bb\Ao / \Ho $). 
$\mA = \mH = \mHpm = 200 $~GeV and
all SUSY parameters except $\tb$ are fixed (from~\cite{LC-PHSM-2003-064}).} 
\label{fig:tanbeta}
\end{figure}

At tree level, the MSSM Higgs sector only depends on $\tb$ and $\mA$. Thus,
if $\mA$ would be measured, $\tb$ could in principle be uniquely determined from the
observed Higgs properties. In particular, the coupling of $\Ao$ to
down-type fermions is directly proportional to $\tb$. Therefore this
coupling which appears in the rate of the $\ee\ra\bb\Ao$ and
$\ee\to\Ao\Ho\to\bb\bb$ processes, as well as
in the total decay width $\Gamma_{\rm A}$ can be used to extract $\tb$
in principle. This has been studied in~\cite{LC-PHSM-2003-064}. 
Due to the large radiative corrections the predictions for the 
observables also depend on other SUSY parameters 
(in particular the sfermion masses and mixings)
which are fixed in this analysis. Therefore the resulting errors 
(see~\ref{fig:tanbeta})
are only valid if all other SUSY parameters,
were precisely known.

A different approach to $\tb$ determination has been proposed 
in~\cite{Beccaria:2002vd}. In a scenario where all SUSY particles are
light compared to the center-of-mass energy, the dependence of the
cross-section for charged Higgs production on $\sqrts$ in the 1 TeV domain 
can be compared
to the logarithmic Sudakov expansion of the cross-section. In particular,
it has been shown, that the first coefficient of the expansion depends only 
on $\tb$.

\begin{figure}[htb]
\centering
\epsfig{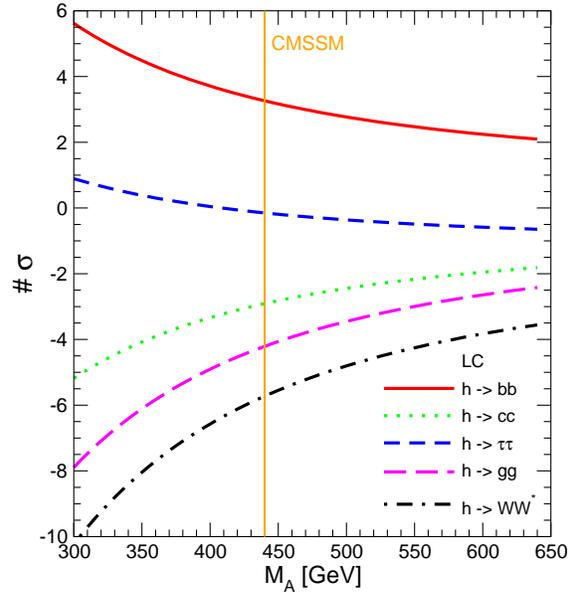}
\caption{Deviation of decay branching ratios of the lightest CP even
Higgs in the constrained MSSM with non-universal Higgs mass (NUHM)
(for $\tb = 10, A_0 = 0, m_{1/2} = 300 $ GeV and $m_0 = 0$)
from their SM values in terms of standard deviations
of the prospective measurement error at the LC as a function of $\tb $.
The dependence on $\mu $ is weak (from~\cite{LC-TH-2002-013}). 
The errors are taken from~\cite{TDR}.}
\label{fig:cmssmsensi}
\end{figure}

A complete study of SUSY parameter determination in the full
MSSM is only possible
when studies of the Higgs sector are combined with information on
sparticle production. Within more constrained SUSY models which assume specific
SUSY breaking schemes Higgs 
observables alone can lead to significant constraints~\cite{Dedes:2003cg}.
As an example, the NUHM (non-universal Higgs mass) 
model has been considered in~\cite{LC-TH-2002-013}.
The NUHM model assumes unification of sfermion masses
and mixing terms as well as unification of gaugino 
mass terms at a high scale. However, in contrast to the mSUGRA (minimal
supergravity) model,
both $\mu$ and $\mA$ are free parameters. In Fig.~\ref{fig:cmssmsensi},
the deviation of branching ratios of the lightest Higgs boson from
the SM is shown for the NUHM scenario as a function of $\mA$.
The deviation is plotted in 
terms of standard deviations
of the prospective measurement error at the LC as described in the TDR.
It can be seen that in particular $\lh\to\bb$ and $\lh\to\WW$ provide
good sensitivity to $\mA$ while the dependence on $\mu$ is only weak.
As a caveat, the values of $\tb$ as well as the other model parameters
are fixed in this study and thus have to be allowed to vary freely in
the study or assumed to
be known from elsewhere in order to translate the plotted deviations
into expected errors on the parameter measurements.

\begin{figure}[htb]
\centering
\epsfig{width=0.82\linewidth,file=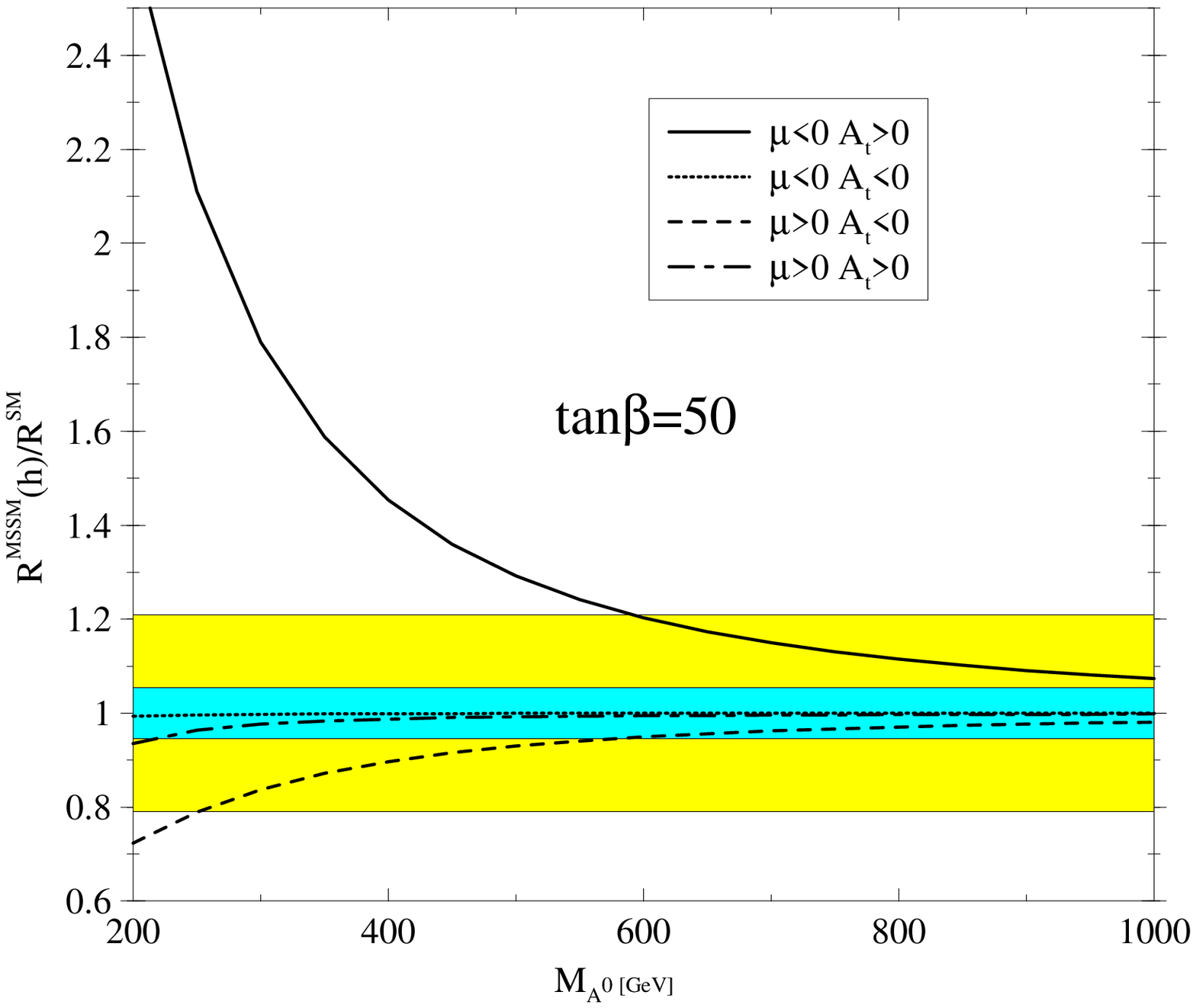}\\
\epsfig{width=0.82\linewidth,file=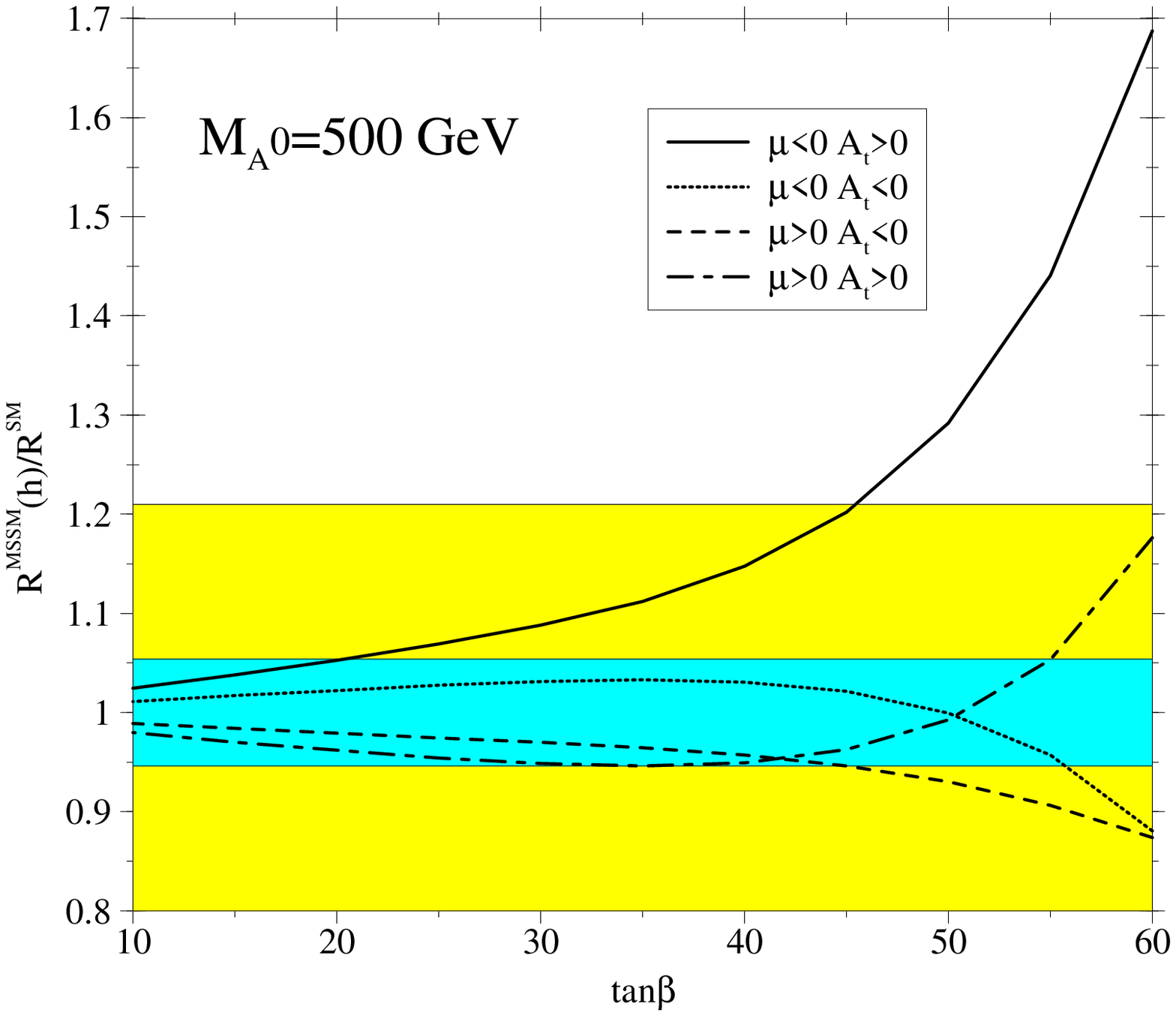}
\caption{Deviation of the ratio ${\cal R} = BR(\lh\to\bb) / BR(\lh\to\tautau)$
as a function of $\mA$ (top) and $\tb$ (bottom). 
(from~\cite{LC-TH-2003-043}. 
The inner band represents the expected measurement error taken from~\cite{TDR}.}
\label{fig:bbttsensi}
\end{figure}

Another study utilizes the ratio 
${\cal R} = BR(\lh\to\bb) / BR(\lh\to\tautau)$~\cite{LC-TH-2003-043}.
At tree level, in the MSSM, this ratio is constant since both b quarks
and $\tau$ leptons are down-type fermions, coupling proportionally to
$\sin\alpha / \cos\beta$ to the $\lh$. A precise measurement of this
ratio is therefore sensitive to the difference of the radiative corrections
to these two decays. In particular at large $\tan\beta$ these corrections
become relevant, allowing to gain sensitivity to the value of $\tan\beta$
itself if all other SUSY parameters are fixed.
The ratio of ${\cal R}^{\mathrm{MSSM}}/{\cal R}^{\mathrm{SM}} $ 
as a function of $\tan\beta$ is shown in Fig.~\ref{fig:bbttsensi}.

\subsection{CP violation in the SUSY Higgs Sector}

In the MSSM the Higgs potential is invariant
under the CP transformation at tree level. However, it is possible to break CP
symmetry in the Higgs sector by radiative corrections, especially by
contributions from third generation 
scalar-quarks~\cite{Pilaftsis:1999qt, Carena:2000yi, LC-TH-2002-016}. 
Such a scenario is theoretically attractive since it provides a 
possible solution to the cosmic
baryon asymmetry~\cite{Carena:2000id}.
In a CP violating scenario the three neutral Higgs bosons,
H$_1$, H$_2$, H$_3$, 
are mixtures of the CP even and CP odd Higgs fields. Consequently, they
all couple to the Z boson and to each other. These
couplings may be very different from those of the CP conserving case.
In the CP violating scenario
the Higgs-strahlung processes $\ee\ra\Hi\Z$ ($i=1,2,3$) and 
pair production processes
$\ee\ra\Hi\Hj$ ($i\neq j$) may all occur, with widely varying cross-sections.

In a case study, for $\mHpm = 200 $ GeV and $\tb = 3$, the sensitivity
of the observable Higgs masses $m_{H_1}$, $m_{H_2}$ and of the 
observed cross-section for $\ee\ra H_1 H_2\ra\bb\bb$ to the real and
imaginary part of the trilinear coupling $A_t$ has been analyzed. 
Under the assumption that the other SUSY parameters are known, the
complex phase of $A_t$ may be extracted from these observables~\cite{klimk}. Clearly,
further studies are needed in order to extract CP-violating SUSY parameters
from the Higgs sector.

\begin{figure}[htb]
\centering
\epsfig{width=0.95\linewidth,file=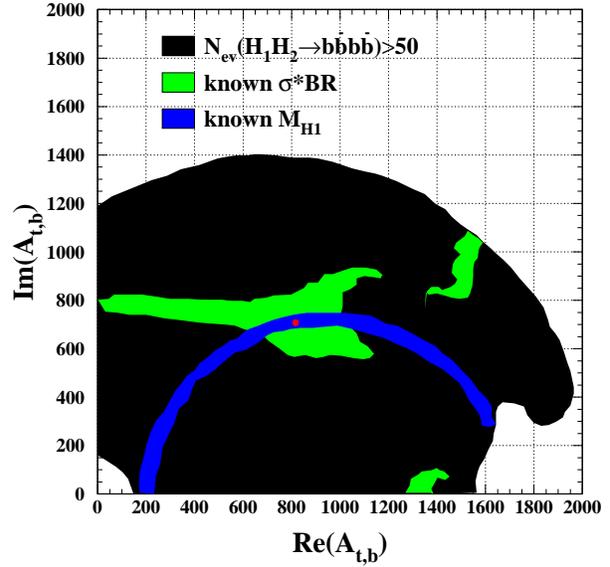}
\caption{Plane of the real and imaginary part of the trilinear coupling
$A_t$ in a CP violating MSSM scenario. In the black region, the Higgs pair 
production process $\ee\to H_1 H_2$ is observable at $\sqrt{s} = 500 $ GeV
with 500 $\fb$. The chosen model point is (750,800) GeV for the (real,imaginary)
part of $A_t$. The dark grey
band is the region which is consistent with the measured lightest Higgs
mass, the medium grey region is consistent with the measured 
$\ee\to H_1 H_2\to\bb\bb$ rate. The real and imaginary part of $A_t$ can
thus be constrained to the overlapping region. Parameters are 
$\mHpm = 200 $ GeV and $\tan\beta = 3$
(from~\cite{klimk}).}
\label{fig:cpvscan}
\end{figure}

\section{EXTENDED MODELS}

\subsection{Genuine Dimension-Six Higgs Operators}

If a light Higgs boson is discovered at the LHC but no additional
particles are seen at the LHC or the LC, it is important to
search for small deviations of the Higgs boson potential from the
SM predictions to probe new physics scales.
If the reason for such small deviations is beyond-SM 
physics at large scales $\Lambda$, the effective operator approach
can be chosen to parameterize the low-energy behavior of such models.
Recently, operators of dimension six have been studied, which involve only 
the Higgs field and which are not severely constrained by precision
electro-weak data~\cite{LC-TH-2003-035}. These operators are
\begin{eqnarray}
    {\mathcal{O}}_1 ={1\over 2}
        \partial_\mu(\Phi^\dagger\Phi) \partial^\mu(\Phi^\dagger\Phi)\quad 
    {\rm and}\quad
        {\mathcal{O}}_2 = -{1\over 3} (\Phi^\dagger\Phi)^3,
    \label{dim6text}
\end{eqnarray}
which lead to a Lagrangian
\begin{equation}
    {\mathcal{L}}' = \sum_i^2 {a_i \over v^2}{\mathcal{O}}_i .
    \label{lp}
\end{equation}
In~\cite{LC-TH-2003-035}, it has been shown that the parameter $a_1$ can be
measured to an accuracy of $0.005 (0.003)$ corresponding to a scale
$\Lambda \approx 4 $ TeV, from 1 $\ab$ of data 
at 500 (800) GeV through the measurement of the production cross-sections
from Higgs-strahlung and WW/ZZ-fusion for $\mH = 120 $ GeV. 
The parameter $a_2$ modifies 
the form of the Higgs potential and thus the Higgs pair production 
cross-section. With the same integrated luminosity, for $\mH = 120 $ GeV,
$a_2$ can be measured to $0.13 (0.07)$ at 500 (800) GeV corresponding to 
a scale $\Lambda \approx $ 1 TeV.

\subsection{Two Higgs Doublet Models}

\begin{figure}[htb]
\centerline{
\epsfig{width=0.49\linewidth,file=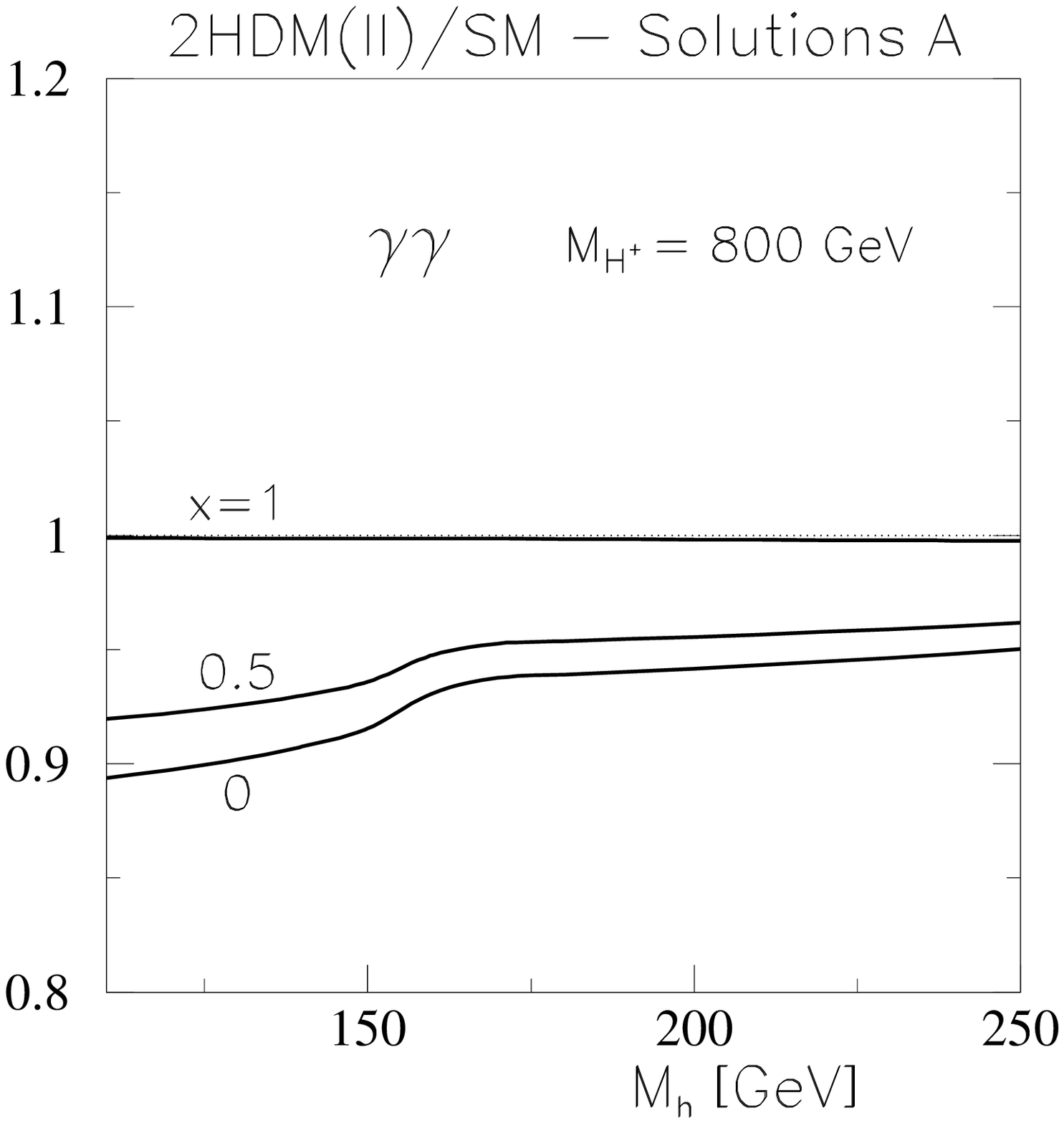}
\epsfig{width=0.49\linewidth,file=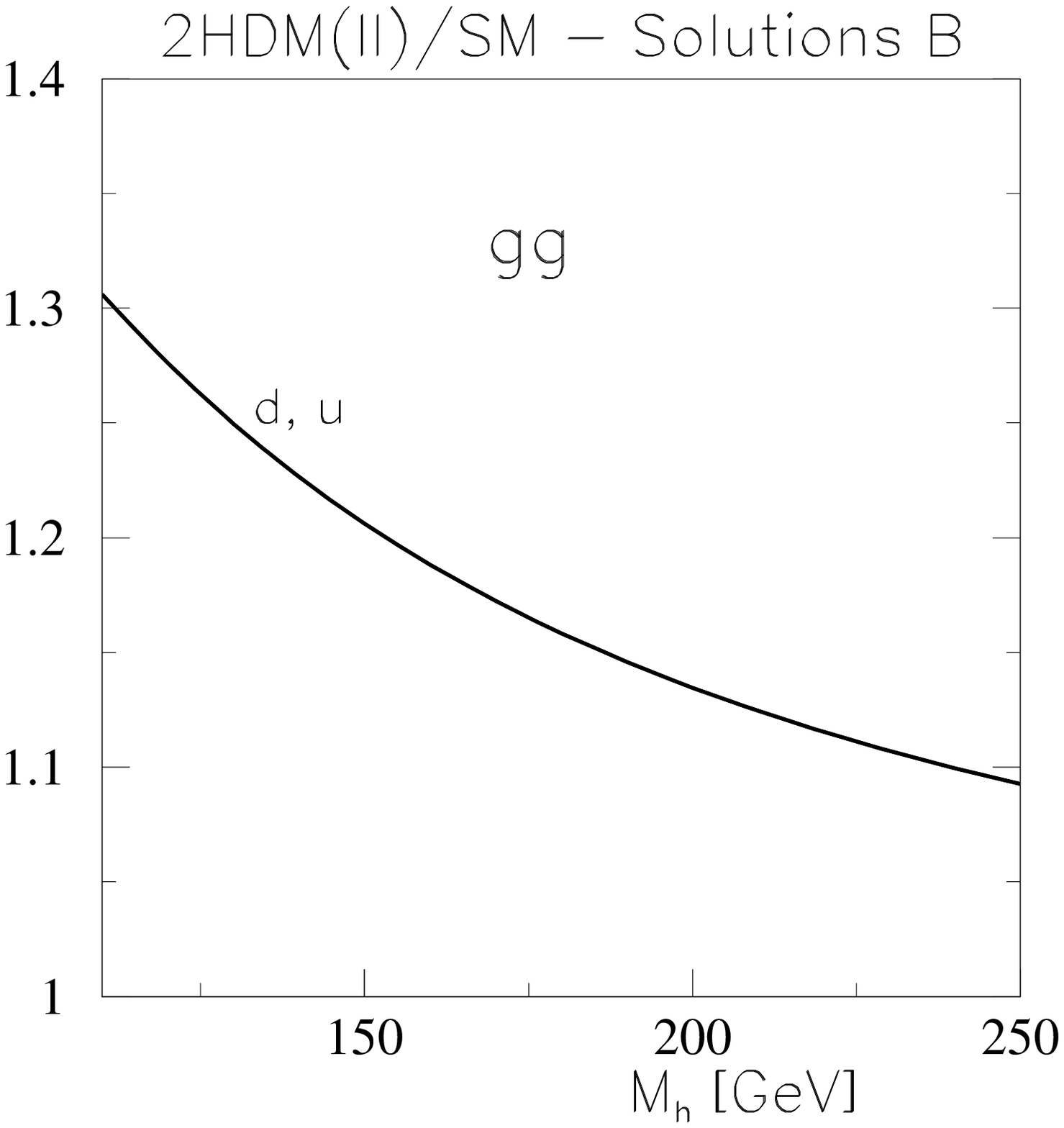}}
\caption{Higgs boson decay
widths in the SM-like 2HDM~(II) relative to the SM decay widths 
as functions of $M_h$.
Left: $h\to\gamma\gamma$ decay widths for a 2HDM scenario with all tree-level
couplings as in the SM up to an overall sign for $M_{H^\pm}=800~\mathrm{GeV}$ 
and $\mu/\sqrt{2}=xM_{H^\pm}$.
Right: $h\to gg$, for a 2HDM scenario with absolute values of
tree level couplings as in the SM but opposite relative sign between 
up-type and down-type fermions (from ~\cite{LC-TH-2003-037}).}
\label{fig:2hdm}
\end{figure}

The prospects for the exploration of general Two Higgs Doublet Models (2HDM)
at a LC have been discussed e.g.~in~\cite{TDR}. During the workshop, a
2HDM scenario has been discussed in which the lightest CP-even Higgs boson
has absolute values of the tree level couplings to fermions and massive
gauge bosons exactly as in the SM and the other Higgs bosons
are heavy (${\cal O}(\mathrm{TeV})$)~\cite{LC-TH-2003-037}. 
Within the 2HDM such a scenario can
be realized differently from the SM in two ways: 
(A) the tree level couplings have the same sign as in the SM or (B)
either up-type or down-type fermions have opposite sign couplings as in the SM.
The only possibility to distinguish such a scenario from the SM is
through loop-induced processes, in particular through the loop-induced
$\gamma\gamma\lh$ and $\mathrm{gg}\lh$ couplings. Depending on $\mlh$ the 
effect can be large enough to be distinguishable from the SM at the LC
(and LHC) from Higgs branching ratio measurements or at a photon collider through
the $\gamma\gamma\to\lh$ process (see Fig.~\ref{fig:2hdm}).

\subsection{NMSSM}

\begin{figure}[htb]
\centerline{
\epsfig{width=0.8\linewidth,file=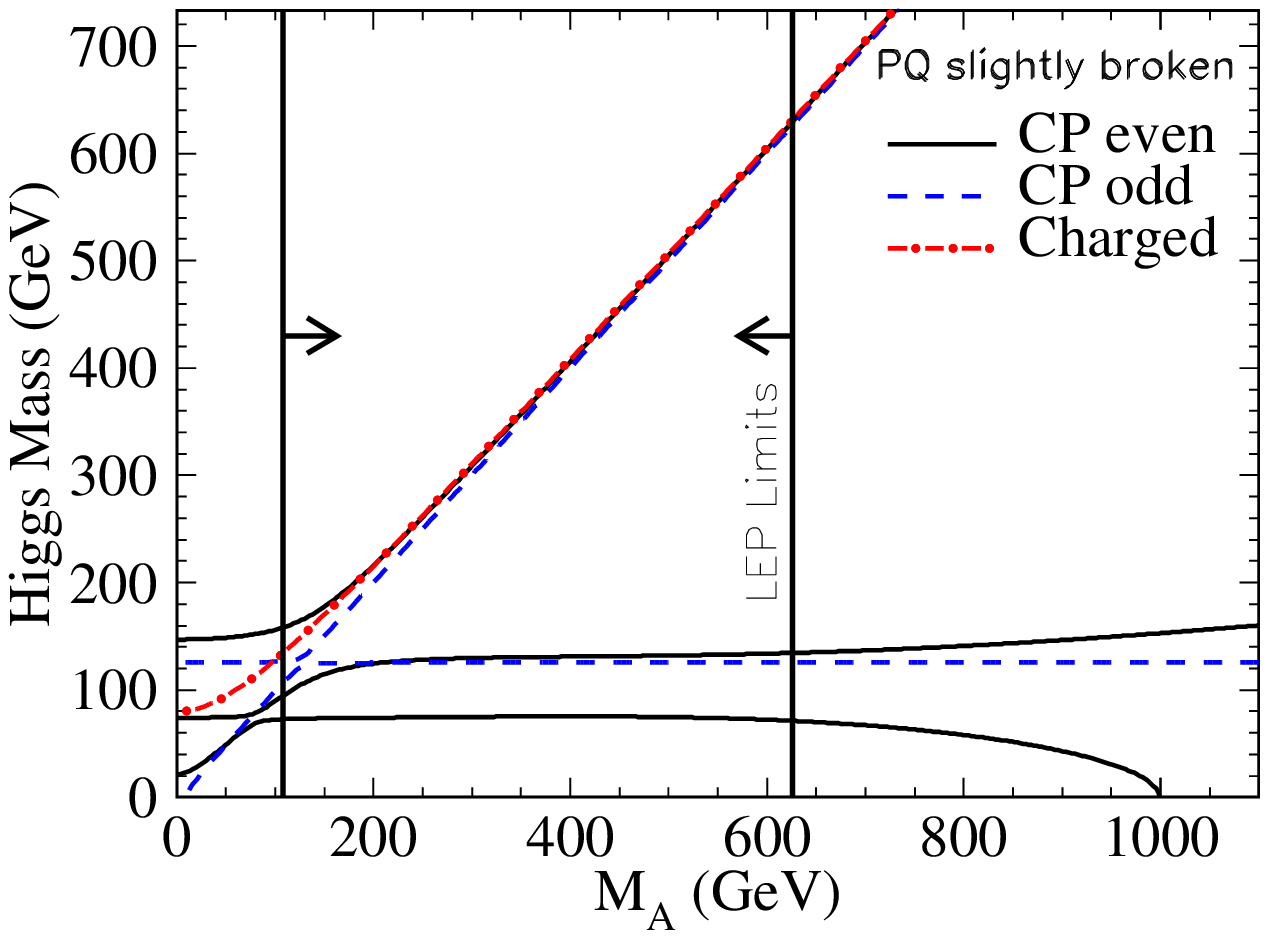}}
\centerline{\epsfig{width=0.8\linewidth,file=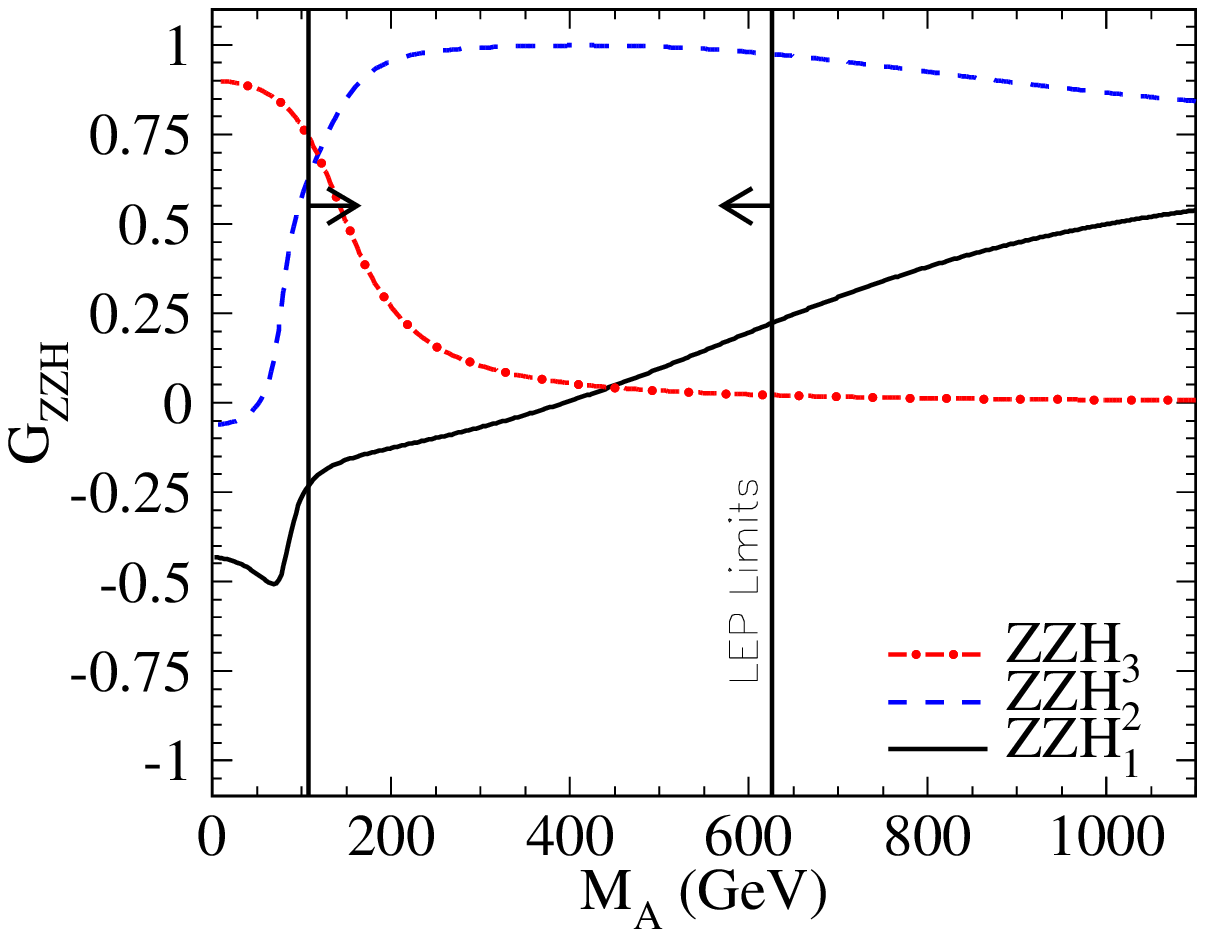}}
\caption{NMSSM Higgs boson properties: masses (upper plot) and couplings
of the CP even Higgs bosons to the 
$\Z$(lower plot)~(from~\cite{LC-TH-2003-034})
for a scenario with slightly broken Peccei-Quinn symmetry
(for $\lambda=0.05$, $\kappa=0.02$,
$v_s=15\, v$, $\tan \beta=3$ and $A_{\kappa}=-100$ GeV). The arrows
denote the region allowed by LEP searches with 95\% confidence.}
\label{fig:nmssm}
\end{figure}

The addition of a Higgs singlet field defines the Next-to-minimal MSSM 
(NMSSM). This addition is theoretically motivated mainly since it
allows a naturally small $\mu$ parameter.
If the associated  
Peccei-Quinn symmetry were unbroken, it would
lead to a massless CP odd Higgs boson which is ruled out. The LC phenomenology
of the model depends on how strong this symmetry is broken. 
The Higgs spectrum of the NMSSM consists of three CP-even and two CP-odd neutral
Higgs bosons and two charged Higgs bosons. The complete LC
phenomenology has recently been reviewed in~\cite{LC-TH-2003-034}.
As an example, the masses of the neutral and charged Higgs bosons and
the coupling of the CP-even Higgs bosons to the Z are shown in 
Fig.~\ref{fig:nmssm} as a function of $\mA$ (defined as the top left parameter
of the CP-odd Higgs mixing matrix, see~\cite{LC-TH-2003-034}). It can be seen
that in a large portion of the parameter space, all three CP-even Higgs
bosons would have significant couplings to the Z, thus significant 
Higgs-strahlung cross-sections at the LC.

\subsection{Higgs Bosons and Extra Dimensions}

Models which postulate the existence of additional space dimensions 
in order to explain the hierarchy between the electro-weak and the
Planck scale have been discussed extensively in recent 
years. Their common feature is that the apparent weakness 
of gravity in our 4-dimensional world is a result of its dilution in the extra 
dimensions. Two scenarios, that of large extra dimensions (ADD)~\cite{add}
and that of warped extra dimensions (RS)~\cite{rs}
have been discussed in particular. The 'classic' signatures involve
deviations of SM processes like $\ee\to\ffbar$ and $\ee\to\WW$
from the virtual exchange of towers of (ADD)~\cite{addhewett}
or single~\cite{rshewett} Kaluza-Klein (KK) excitations of gravitons, or
their real emission together with SM fermions or gauge bosons~\cite{gravemis}.
These modes have been studied experimentally e.g.~in the TESLA TDR.

More recently, also the impact of extra dimensions on the Higgs boson
phenomenology has been studied. In the ADD scenario, two effects have been
analyzed: 

1. A modification of the quasi-resonant $\WW\to\Ho$ production process
through interference of the SM amplitude
with the imaginary part of the graviton/graviscalar KK exchange 
amplitude~\cite{LC-TH-2003-011}. In order to
yield a significant modification, a large total Higgs width is needed 
(i.e.~large $\mH$), which implies on the other hand a large center-of-mass
energy. While the graviscalar contribution only modifies the normalization
of the cross-section 
(by few percent for $\sqrts = $ 1 TeV, $\mH = $ 500 GeV and 2 extra 
dimensions at a fundamental Planck scale of 1 TeV),
a significant change of the angular distribution is expected from the
spin-2 graviton exchange. 

2. A modification of the process $\ee\to\Ho\Ho\Z$ and the existence of
the process $\ee\to\Ho\Ho\gamma$ which is absent at 
tree level in the SM~\cite{Deshpande:2003sy}. For a 1 TeV LC and 
$\mH = $ 120 GeV, a sizable correction to $\ee\to\Ho\Ho\Z$ both in
normalization and angular distribution is expected for fundamental Planck
scale up to a few TeV. Furthermore, the cross-section for $\ee\to\Ho\Ho\gamma$
exceeds 0.1 fb for a fundamental Planck scale below approximately 2 TeV.
In~\cite{Deshpande:2003sy}, expected 5 $\sigma $ discovery limits on
the fundamental Planck scale of 880--1560 (1640--2850) GeV have been derived
at $\sqrts = $ 500 (1000) GeV for 6--3 extra dimensions.

\begin{figure}[htb]
\centering
\epsfig{width=0.7\linewidth,file=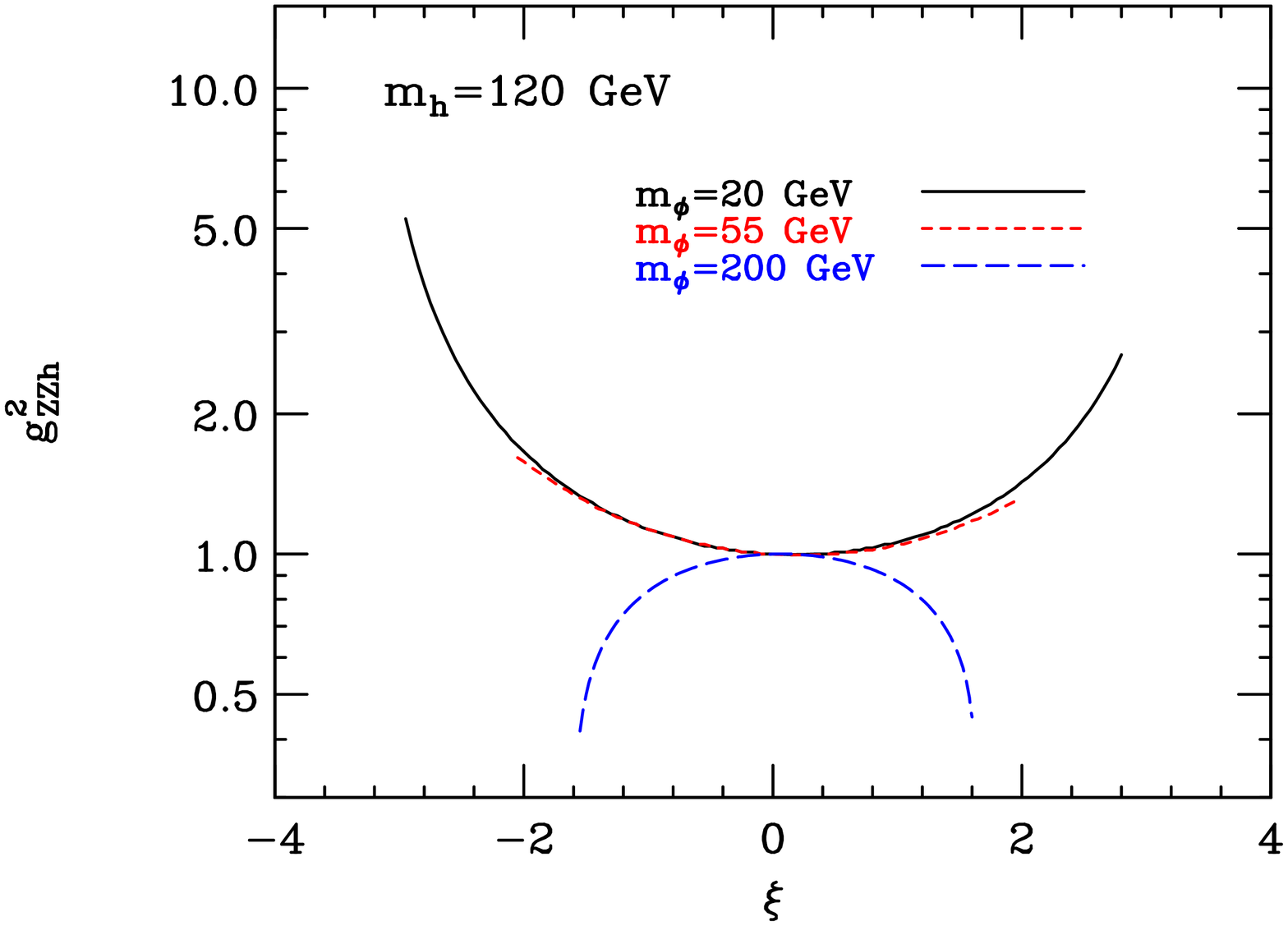,angle=90}\\
\epsfig{width=0.7\linewidth,file=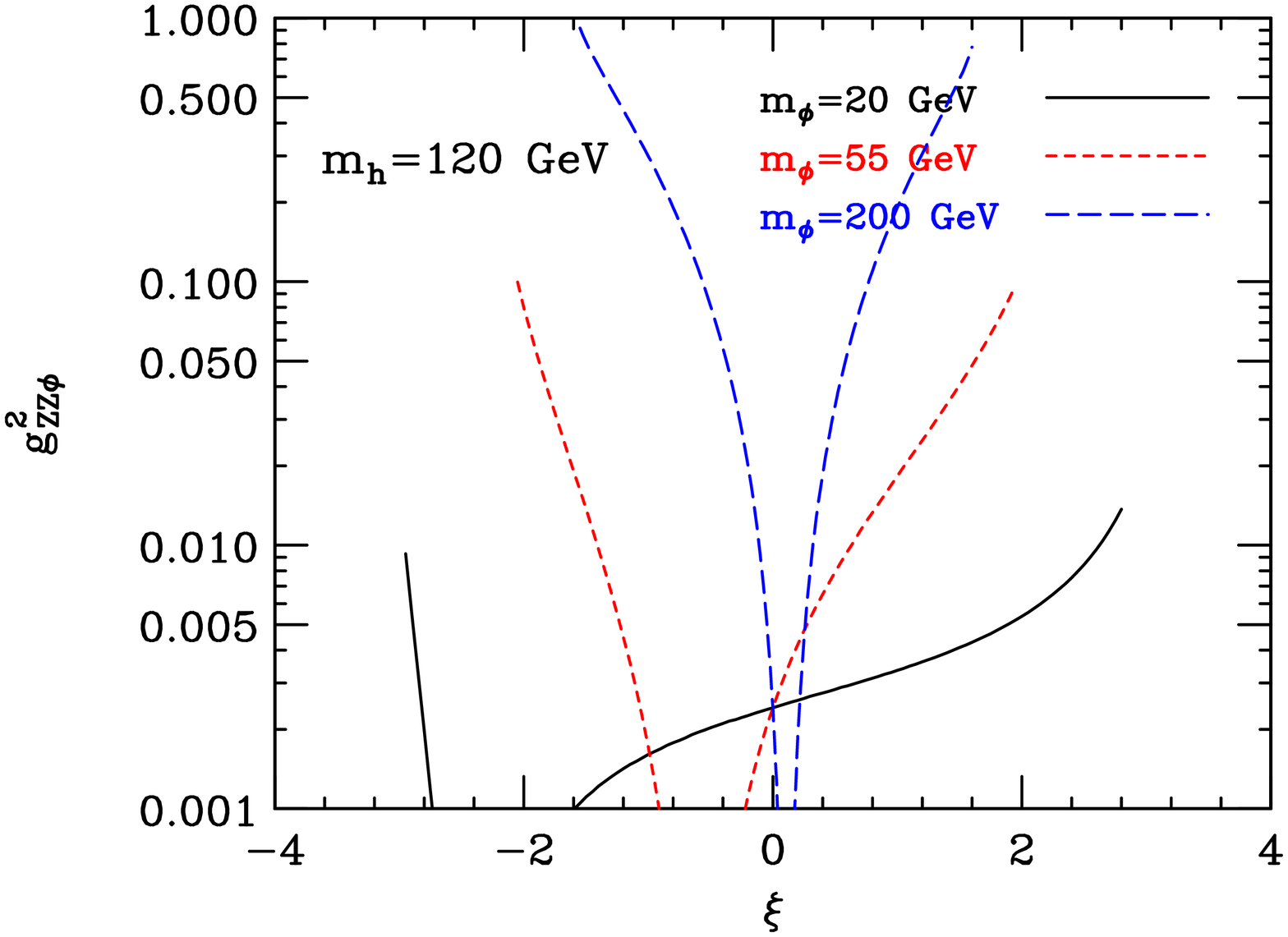,angle=90}
\caption{Effective coupling of the Higgs boson (upper) and the Radion (lower)
to Z boson (from \cite{radionrev}).}
\label{fig:radionprop}
\end{figure}

In the RS scenario, the influence on the Higgs sector might be much 
more drastic. Besides the spin-2 KK graviton excitations, graviscalar 
excitations, called Radions, are predicted~\cite{radions}. They are
predicted to couple to SM particles through the trace of the energy-momentum
tensor, i.e.~up to the trace anomaly of QCD, very similar to the Higgs boson.
The lightest Radion might in fact be lighter than the lightest graviton
excitation and thus the discovery channel for the model. Higgs boson and
Radion may exhibit kinetic mixing, which leads to a 
modification of both Higgs boson and Radion properties, in particular their
couplings to gauge bosons and fermions. For a review of the Radion 
phenomenology, see e.g.~\cite{radionrev}. The Radion sector is governed
by 3 parameters: the strength of the Radion-matter interactions described
by an energy scale $\Lambda_\phi$, the mass of physical Radion, $m_\phi$,
and the Radion-Higgs mixing parameter $\xi$. In Fig.~\ref{fig:radionprop},
the effective couplings squared of the Higgs boson and the Radion (relative
to those of a SM Higgs boson) are shown
for the choice $\Lambda_\phi = 5 $ TeV, and three values of the Radion mass
(20, 55, 200 GeV) as a function of $\xi$. Large deviations of the Higgs
couplings from their SM values are expected if there is large Radion Higgs mixing present. The Radion itself has couplings which are reduced by a factor
$v/\Lambda_\phi$ with respect to those of a SM Higgs in the case of no mixing,
which requires high luminosity for direct discovery. The sensitivity of the
trilinear Higgs coupling to Radion admixtures 
has been studied as well in~\cite{radionrev}.

\begin{figure}[htb]
\centering
\epsfig{width=\linewidth,file=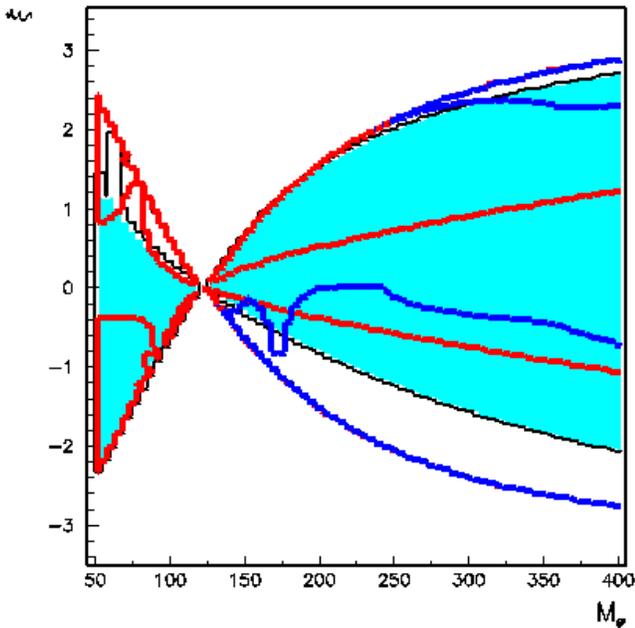}
\caption{Sensitivity to Radions at LHC and LC: parameter plane
of the Radion Higgs mixing angle $\xi$ and the Radion mass $M_\phi$
for $\mH = 120 $ GeV. In the shaded regions, the LHC can observe 
the Higgs boson in the $gg\to\Ho\to\gaga$ channel with 30 fb$^{-1}$ in
one experiment. The dark (blue) lines indicate the regions 
where LHC can observe the Radion in the $gg\to\Phi\to 4\ell$ channel.
The grey (red) lines indicate the regions where at the LC a $>2.5\sigma$
deviation of the Higgs branching ratio BR($\Ho\to\bb$) is observable.
For more details see~\cite{radionlhclc}.}
\label{fig:battaglia}
\end{figure}

The LC capability of precisely measuring the Higgs branching ratios 
$\Ho\to\bb$ and $\Ho\to\WW$ has been exploited in~\cite{radionlhclc}. 
In Fig.~\ref{fig:battaglia}, the regions where the LC would observe
larger than 2.5$\sigma$ deviations of the Higgs branching ratios due to
Radion Higgs mixing is shown together with the regions where the LHC
can observe the Higgs bosons. In particular the regions in which the
LHC might be blind to the Higgs boson are well covered by the LC.
A study of the sensitivity of the WW-fusion channel to Radion effects
has also been presented at the workshop~\cite{Datta:2003kf}.

\section{RELATION TO THE LHC}

A Higgs boson with SM-like properties will most likely be discovered
at the Large Hadron Collider LHC. In recent years, the potential of
the LHC to make measurements of Higgs boson properties has been investigated.
For a recent summary of the ATLAS studies, see~\cite{duehrssen}.
In most cases the capabilities of a LC are superior to those of the
LHC as far as Higgs physics is concerned. In particular, \it no model-independent
measurements \rm of Higgs boson couplings are possible at the LHC. 
However, there are cases where the synergy of both colliders is vital
and rewarding. Examples are in the determination of the top Yukawa coupling, in the
mass reach for heavy SUSY Higgs bosons, and in LHC measurements on
third generation squark properties in order to constrain the interpretation
of a supersymmetric Higgs sector. These examples are currently been
worked out in more detail in a world-wide LHC/LC study group~\cite{lhclc}.

\section{SUMMARY AND OUTLOOK}

The precision study of Higgs bosons is at the core of the physics program
of a future linear collider. In the course of the extended ECFA/DESY
study 2001-2003 this physics case has been developed further: 
the precision of theoretical calculations has been improved, 
the implication of new theoretical models has been investigated and
the experimental studies of the LC sensitivity have been extended and
improved.

The studies are vital for the preparation of the worldwide LC project
and will be continued both in the three regions America, Asia, and Europe
and in worldwide workshops. In Europe, the study will continue in the
framework of a new ECFA study. Major goals of this new study are to continue
to incorporate new theoretical ideas and to 
improve the precision of theoretical predictions.
On the experimental side, a more detailed
study of systematic limitations, impact of machine conditions and in
particular dependence of the precision on specific detector properties
are of utmost importance.

\section{ACKNOWLEDGMENTS}

I would like to warmly thank all contributors to the Higgs working group
for their huge efforts to make the workshop a success. In particular
the work of my co-convenors M.~Battaglia, A.~Djouadi, E.~Gross, and
B.~Kniehl is greatly acknowledged.

\end{document}